\documentclass[nofootinbib,notitlepage,superscriptaddress,10pt,aps,pra,twocolumn]{revtex4-2}

\usepackage{xcolor}
\usepackage[utf8x]{inputenc}
\usepackage{amsmath}
\usepackage{amssymb}
\usepackage{graphicx}
\usepackage{float}
\usepackage[normalem]{ulem}
\usepackage{epsf,epsfig}
\usepackage{psfrag}
\usepackage{color}
\usepackage{multirow}
\usepackage{diagbox}
\usepackage{url}
\usepackage{longtable}

\makeatletter
\newcommand\stroke[1]{\mathpalette\stroke@aux{#1}}
\def\stroke@aux#1#2{%
  \ooalign{%
    \hfil$#1^{\;\, \_\hspace{-0.05cm}\_}$\hfil\cr
    \hfil$#1#2$\hfil\cr
  }%
}
\makeatother

\begin{document}

\title{Redshift leverage for the search of GRB neutrinos\\ affected by quantum properties of spacetime}

\author{Giovanni Amelino-Camelia}
\affiliation{Dipartimento di Fisica Ettore Pancini, Universit\`a di Napoli “Federico II”, Complesso Univ. Monte S. Angelo, I-80126
Napoli, Italy}
\affiliation{Istituto Nazionale di Fisica Nucleare, Sezione di Napoli, Complesso Univ. Monte S. Angelo, I-80126 Napoli, Italy}
\author{Giacomo D'Amico}
\altaffiliation[Present address: ]{Institut de Física d’Altes Energies (IFAE), The Barcelona Institute of Science and Technology (BIST), E-08193 Bellaterra (Barcelona), Spain}
\affiliation{Department for Physics and Technology, University of Bergen,  NO-5020 Bergen, Norway}
\author{Vittorio D'Esposito}
\affiliation{Dipartimento di Fisica Ettore Pancini, Universit\`a di Napoli “Federico II”, Complesso Univ. Monte S. Angelo, I-80126
Napoli, Italy}
\affiliation{Istituto Nazionale di Fisica Nucleare, Sezione di Napoli, Complesso Univ. Monte S. Angelo, I-80126 Napoli, Italy}
\author{Giuseppe Fabiano}
\affiliation{Physics Division, Lawrence Berkeley National Laboratory, Berkeley, CA 94720, USA}
\affiliation{Department of Physics, University of California, Berkeley, CA 94720, USA}
\affiliation{Centro Ricerche Enrico Fermi, I-00184 Rome, Italy}
\author{Domenico Frattulillo}
\affiliation{Istituto Nazionale di Fisica Nucleare, Sezione di Napoli, Complesso Univ. Monte S. Angelo, I-80126 Napoli, Italy}
\author{Giulia Gubitosi}
\affiliation{Dipartimento di Fisica Ettore Pancini, Universit\`a di Napoli “Federico II”, Complesso Univ. Monte S. Angelo, I-80126
Napoli, Italy}
\affiliation{Istituto Nazionale di Fisica Nucleare, Sezione di Napoli, Complesso Univ. Monte S. Angelo, I-80126 Napoli, Italy}
\author{Dafne Guetta}
\affiliation{Capodimonte Observatory, INAF-Naples, Salita Moiariello 16, Naples 80131, Italy}
\author{Alessandro Moia}
\affiliation{Dipartimento di Fisica Ettore Pancini, Universit\`a di Napoli “Federico II”, Complesso Univ. Monte S. Angelo, I-80126
Napoli, Italy}
\author{Giacomo Rosati}
\affiliation{Institute for Theoretical Physics, University of Wroc{\l}aw, Pl. Maksa Borna 9, Pl–50-204 Wroc{\l}aw, Poland}
\affiliation{Dipartimento di Matematica, Università di Cagliari, via Ospedale 72, 09124 Cagliari, Italy}

\begin{abstract}
Some previous studies based on IceCube neutrinos had found intriguing  preliminary evidence that some of them might be GRB neutrinos with travel times affected by quantum properties of spacetime delaying them proportionally to their energy,
an effect often labeled as ``quantum-spacetime-induced in-vacuo dispersion".
Those previous studies looked for candidate GRB neutrinos in a fixed (neutrino-energy-independent) time window after the GRB onset and relied rather crucially on  crude estimates of the redshift of GRBs whose redshift has not been measured.
We here introduce a complementary approach to the search of quantum-spacetime-affected GRB neutrinos which restricts the analysis to GRBs of sharply known redshift, and, in a way that we argue is synergistic with having sharp information on redshift, 
adopts a neutrino-energy-dependent time window. We find that
knowing the redshift of the GRBs strengthens the analysis enough to compensate for the fact 
that of course the restriction to GRBs of known redshift reduces the number of candidate GRB neutrinos. And rather remarkably our estimate of the magnitude of the in-vacuo-dispersion effects  is fully consistent with what had been found using the previous approach.
Our findings are still inconclusive, since their
 significance  is quantified by a  $p$-value of little less than  $0.01$, 
 but provide motivation for monitoring the accrual of neutrino observations by IceCube and KM3NeT as well as for further refinements of the strategy of analysis here proposed.\newline
\end{abstract}

\maketitle

\section{Introduction and new selection criteria}\label{sec:intro}
The current generation of neutrino telescopes was devised to mark the start of neutrino astrophysics, and indeed IceCube firmly established the observation of cosmological neutrinos~\cite{IceCube:2013low}, but so far not much astrophysics has been done with neutrinos:  there is only evidence of a neutrino signal from the 
active galaxy NGC 1068~\cite{IceCube:2022der} and from
the flaring blazar TXS 0506+056
(see, {\it e.g.}, Refs.~\cite{IceCube:2018cha,Ellis:2018ogq}).
With respect to pre-IceCube expectations what is clearly missing are observations of neutrinos from GRBs (gamma-ray bursts). The prediction of a neutrino emission associated with
GRBs is generic within the most
widely accepted astrophysical models~\cite{Piran:1999bk} 
and pre-IceCube studies estimated about a handful of GRB-neutrino observations per year of operation~\cite{Waxman:1997ti, Meszaros:1997qf, Guetta:2003wi, Ahlers:2011jj}; yet, as IceCube nears 15 years of operation, still not a single GRB neutrino has been reported. The simplest (and most likely) explanation of this huge underperformance of IceCube is that our GRB models must be revised in such a way to produce a much lower neutrino flux, but it is intriguing that in principle our failure to observe GRB neutrinos could also be attributed to some plausible quantum properties of spacetime which had already been investigated for independent reasons in the quantum-gravity literature:
IceCube searches might have failed to find GRB neutrinos 
because they assume  that a GRB neutrino should
be detected in very close temporal coincidence with the
associated gamma rays, but a sizable mismatch between the neutrino detection time and the time of detection of the electromagnetic  GRB signal would be expected in presence of ``in-vacuo dispersion", a much-studied effect such that quantum properties of spacetime slow down\footnote{While the theoretical prejudice of most quantum-spacetime researchers favors effects that slow down particles, there is no conclusive theoretical argument ruling out that quantum-spacetime effects might instead speed up particles. We offer some data-based observations relevant for this issue in parts of Sec.~\ref{sec:refinements}.} particles proportionally to their energies (see, {\it e.g.}, Refs.~\cite{grbgac,jacobpiran, gacsmolin, gampul, urrutia, gacmaj, myePRL, Mattingly:2005re, GAClivingReview, COSTreview} and references therein).

We here focus exclusively on the most studied scenario for in-vacuo dispersion which for our purposes is conveniently characterized in terms of the following relationship~\cite{Ryan,RyanLensing,NatureAstro,MaNeutriniFirst,HuangLiMaNeutriniNearTeV}: 
\begin{equation}
\Delta t = \eta D(z) \frac{E}{M_{P}}   \, ,
\label{mainnewone}
\end{equation}
where $\Delta t$ is the
contribution to the travel time of the neutrino due to
quantum-gravity-induced in-vacuo dispersion, $E$ is the neutrino energy,
 $M_{P}$ denotes
the Planck scale ($\sim 10^{16}$~TeV), and
$D(z)$ is a function of the redshift $z$ of the source,
$$D(z) = \int_0^z d\zeta \frac{(1+\zeta)}{H_0\sqrt{\Omega_\Lambda + (1+\zeta)^3 \Omega_m}} \,$$
($\Omega_\Lambda$, $H_0$ and $\Omega_m$ denote, as usual,
respectively the cosmological constant, the Hubble parameter and the matter fraction, for which we take the values given in Ref.~\cite{Planck:2018vyg}).

We restrict to the scenario with purely systematic effects, in which $\eta$ is simply a parameter to be determined experimentally. When a larger collection of astrophysical neutrinos becomes available (and therefore analyses such as ours can attempt multi-parameter investigations), it will be possible to consider the more general case of a mixture of systematic and fuzzy effects~\cite{GAClivingReview, COSTreview, gacPIRANnaturephysics2015}, in which $\eta$ is described as a random variable (taking different values for different neutrinos) with mean $\eta_{0}$ and dispersion $\delta\eta$.
We also postpone the extension of our phenomenological model to allow for the possibility that there might be intrinsic at-the-source effects causing a sizable offset between the typical neutrino emission time and the GRB onset. A first step for modeling the impact of such intrinsic effects on in-vacuo dispersion analyses would be to add to Eq.(\ref{mainnewone}) an extra term of the form ${\bar{\tau}} (1+z)$, where $\bar{\tau}$ is the average neutrino emission time offset with respect to the GRB onset and the factor $(1+z)$ accounts for ordinary time dilation, but investigations of such a refined model need more astrophysical neutrinos than presently available.

The effects of 
Eq.~(\ref{mainnewone})
are totally unnoticeable on terrestrial scales, but for the observation of distant astrophysical sources the factor $D(z)$ can be large enough to compensate for the Planck-scale suppression~\cite{grbgac,jacobpiran, gacsmolin, gampul, urrutia, gacmaj, myePRL, Mattingly:2005re, GAClivingReview, COSTreview}.

Indeed, in-vacuo dispersion for photons has been investigated very intensely over the last 15 years (see, {\it e.g.}, Refs. \cite{GAClivingReview, COSTreview} and references therein), with some studies even reaching sensitivity of the order of the Planck scale (see, {\it e.g.}, Ref.~\cite{FermiGBMLAT:2009nfe}). It is difficult to say exactly which values of $\eta$ have been excluded for photons, since the magnitude of in-vacuo-dispersion effects could be comparable to known (but poorly understood) intrinsic spectral lags and a delayed emission of the relevant photons cannot be ruled out (some of the most energetic photons from a GRB could be emitted within the long afterglow phase). Still, a balanced perspective is that, for photons, we are close to Planck-scale sensitivity~\cite{GAClivingReview, COSTreview}.

The investigation of in-vacuo dispersion for neutrinos has been lagging far behind: the current bound on in-vacuo dispersion for neutrinos is still based on the historic observation of neutrinos from the SN1987a supernova, and only amounts to $\eta \lesssim 10^9$~\cite{Ellis:2008fc}. This is very unsatisfactory, especially in light of the fact that the quantum-spacetime literature provides  motivation for studying in-vacuo dispersion for photons and neutrinos separately. In particular, quantum-spacetime models based on Planck-scale discretization predict in-vacuo dispersion for neutrinos but not for photons~\cite{Alfaro:1999wd,Sahlmann:2002qj}, and this is consistent with results from models based on causal sets~\cite{Philpott:2010cm}. Moreover, within a popular effective-theory approach to quantum-spacetime  properties one describes in-vacuo dispersion for neutrinos in a way that is completely independent from that applicable to photons~\cite{Myers:2003fd}. Given the typical energies of the astrophysical neutrinos observed by current telescopes, even just  a single observation of a GRB neutrino would allow us to constrain in-vacuo dispersion for neutrinos rather tightly. However, as mentioned at the beginning, no such observation has been reported yet by studies assuming no in-vacuo dispersion.

As observed in Refs.~\cite{Ryan,RyanLensing,NatureAstro,MaNeutriniFirst,HuangLiMaNeutriniNearTeV, gacGuettaPiran, natastro2023} the search of GRB neutrinos affected by in-vacuo dispersion must necessarily rely on a statistical approach: we could never be sure that a certain neutrino is a GRB neutrino, but we might at some point observe enough GRB-neutrino candidates (neutrinos that could come from a GRB, assuming in-vacuo dispersion) to be sure that at least some of them are actually GRB neutrinos affected by in-vacuo dispersion.
This is due to the size of the time window that one must adopt in order to investigate the hypothesis of in-vacuo dispersion. Standard searches of GRB neutrinos (assuming $\eta =0$, no in-vacuo dispersion) can assume that the neutrino and the electromagnetic signal from the GRB travel essentially at the same speed, and therefore their GRB-neutrino candidates are looked for within a tiny time window, rendering background issues insignificant. For $\eta \neq 0$, also taking into account that the neutrino energy will have some uncertainty (and however one will inevitably always end up testing an hypothesis of $\eta$ taking values in a certain range), the difference in observation time between the neutrino and the electromagnetic signal would be sizably uncertain, and  the time window for searches of GRB-neutrino candidates would have to be correspondingly large, resulting in background issues such that one could not be sure that a specific GRB-neutrino association is correct.

Evidently the effectiveness of such a statistical approach depends rather crucially on the criteria used for the selection of candidate GRB neutrinos. And our main objective here is to test a novel proposal for these criteria. Testing alternative strategies of analysis on presently-available IceCube data is also important because it can set the stage for a later, more mature, phase of the research program, eventually using also data from KM3Net~\cite{km3net} and IceCube-Gen2~\cite{icecubegen2}.

Previous searches~\cite{Ryan,RyanLensing,NatureAstro,MaNeutriniFirst,HuangLiMaNeutriniNearTeV, gacGuettaPiran, natastro2023} of GRB neutrinos affected by in-vacuo dispersion
looked for neutrinos within a time window of fixed size
(neutrino-energy-independent)  after the GRB onset, and put on the same footing both
GRBs whose redshift has not been measured
(then having to estimate that redshift crudely, since the conjectured effect is redshift dependent) and
GRBs of known redshift.
Including GRBs whose redshift has not been measured has the advantage of a larger number of candidate GRB neutrinos, but of course renders the analysis vulnerable to the assumptions made to roughly estimate the redshifts. Moreover, the fixed-size time window, while  easily handled computationally, imposes a restriction of the analysis to a corresponding limited range of neutrino energies: since the sought effect grows linearly with energy, a time window adapted to energies of, say, 100 TeV will inevitably be too small for neutrinos with energy much greater than 100 TeV, and in an appropriate sense it would also be too large a time window for neutrinos of energy smaller than 100 TeV (when the time window is much wider than the one really needed by the sought effect, the analysis ends up being dominated by the background).
In spite of these limitations, Ref.~\cite{natastro2023}, the latest such study (of which some of us were authors), provided the estimate $\eta = 21.7 \pm 4.5$ with a $p$-value\footnote{As will be clearer for our readers when, later in this manuscript, we derive an analogous $p$-value for our novel selection criteria, this $p$-value, which in Ref.~\cite{natastro2023} was called false alarm probability, is the probability to find at least as many (and as good) GRB-neutrino candidates as were actually found in the data under the null hypothesis of no in-vacuo dispersion ($\eta=0$), whereas $\eta = 21.7 \pm 4.5$ is an estimate of $\eta$ obtained from the actually found GRB-neutrino candidates.}  $P_{[34]}=0.007$.

We here take the results of Ref.~\cite{natastro2023} as a starting point for testing a novel strategy of analysis. Rather than adopting a fixed-size time window and correspondingly restricting the neutrino-energy range, we consider all the neutrinos in the sample, regardless of the energy, but restrict the GRB catalogue to GRBs of known redshift. Any GRB and neutrino times of arrival are then regarded as compatible if they satisfy Eq.~(\ref{mainnewone}) for some $\eta\in\left[12.7,30.7\right]$ (the two-standard-deviation interval obtained from the estimate $\eta = 21.7 \pm 4.5$ found in Ref.~\cite{natastro2023}) within two standard uncertainties in the neutrino energy. 
As here shown in Sec.~\ref{main}, the fact that, by relying exclusively on GRBs of known redshift, we find fewer GRB-neutrino candidates is more than compensated, for what concerns the overall statistical significance, by the sharper setup of the analysis which is available when the redshift of the GRBs is known.

Also for what concerns the assessment of the directional compatibility between a neutrino and a GRB the criteria here adopted differ from those of previous analogous studies. This change, however, does not reflect a strategic choice: it is rather due to a qualitative upgrade in the neutrino directional information provided by IceCube. Previous IceCube data releases provided directional information in terms of approximate gaussian uncertainties, whereas the latest data release characterizes the uncertainty in the direction of a neutrino in terms of an 8-parameter Fisher-Bingham (FB8) distribution on the sphere~\cite{Yuan:2019dxx}. The asymmetry and complexity of this distribution makes it impossible to reduce the assessment of its directional compatibility with the directional PDF of a known-redshift GRB to a straightforward comparison of their most probable values, as done, {\it e.g.}, in~\cite{natastro2023}. Rather, we follow other studies that handled similarly complex directional uncertainties (see, {\it e.g.}, Ref.~\cite{LIGOScientific:2017zic}) and rely on the value of the statistic $\mathcal{S}=\int P_{\nu}(\Omega)P_{GRB}(\Omega)d\Omega$, where $P_{\nu}(\Omega)$ and $P_{GRB}(\Omega)$ are the angular distributions of the neutrino and the GRB, respectively. This $\mathcal{S}$ can be computed for angular distributions of arbitrary shape and is a good measure~\cite{LIGOScientific:2017zic} of angular compatibility (in particular, its value increases as the peaks of the two distributions grow closer). We thus regard the directions of a GRB-neutrino pair as compatible if the corresponding $\mathcal{S}$ satisfies $\mathcal{S}\geq\mathcal{S}_{cut}$ for some fixed reference value $\mathcal{S}_{cut}$. In setting up our study we adopted  $\mathcal{S}_{cut} = 1/4\pi$, which is the value taken by $\mathcal{S}$ whenever one of the distributions is the uniform distribution on the sphere. This choice is intuitively natural, as it would be difficult to claim any degree of directional compatibility if two uniform, uninformative distributions would be given a better score. In any case, after performing  our main analysis (here reported in Sec.~\ref{main}), we explored the dependence of our results on the choice of $\mathcal{S}_{cut}$, finding that it is rather weak (see Sec.~\ref{sec:refinements}).

\section{IceCube neutrinos and GRBs of known redshift}\label{sec:data}
Our analysis relies on the latest HESE data release by IceCube  (7 September 2023), available at~\cite{icecubedatarelease}. We chose to work with the HESE sample because it is the IceCube sample least contaminated by atmospheric neutrinos, especially above 50-60 TeV \cite{IceCube:2020wum}. As done in previous analogous studies~\cite{Ryan,RyanLensing,NatureAstro,MaNeutriniFirst,HuangLiMaNeutriniNearTeV, natastro2023}, we restrict to shower (or cascade) events, since track events have very poor energy estimates (and of course the quality of the energy information is crucial for in-vacuo-dispersion studies)~\cite{IceCubeEnergy2013,jackJrEnergy2018}. An additional advantage of working with shower events is that they are even more robust against atmospheric contamination (for the shower morphology, the atmospheric neutrino flux above 1 TeV is suppressed by about an order of magnitude with respect to the track morphology~\cite{IceCube:2023ame}).

A sizable part of the effort we devoted to this study was aimed at compiling a reliable catalogue of GRBs 
with actually measured (and not just estimated or loosely constrained) redshifts. In fact, we found that all already available GRB catalogues either imposed further constraints on the GRB properties (narrower observation window, restriction to specific GRB observatories, GRB type, etc.) or were not sufficiently accurate (misreporting or omitting direction or redshift data, not taking into account the latest relevant GCNs, etc.). The resulting list is reported here in Appendix A. In about 90\% of cases it agrees with an analogous general-purpose catalogue, not restricted to known-redshift GRBs, maintained by the IceCube collaboration~\cite{icecubeGRBs}. The previous study~\cite{natastro2023} relied on this resource, but, after checking one-by-one each of its known-redshift entries, we noticed that some of the information relevant for our analysis was inaccurately reported, probably due to the automated population of the database. In most cases this was quickly corrected according to the web catalogue~\cite{maxplanckGRBs} maintained by Jochen Greiner at the MPE, which we found to be the most accurate and comprehensive among publicly available lists. Checking by hand all the relevant GCNs and consulting additional catalogues of known-redshift GRBs, we were then able to correct the few errors left and also identify a few known-redshift GRBs whose redshift is not reported in Greiner's table (see Appendix \ref{GRBcatalogue} for details).

In the end, all these efforts did not actually pay off in terms of a significant improvement of our analysis: as regards the four GRBs of known redshift playing a crucial role in our main analysis (see Sec.~\ref{main}), our corrected and augmented catalogue agrees with both~\cite{icecubeGRBs} and~\cite{maxplanckGRBs}. Still, we hope that the effort we put in preparing Appendix \ref{GRBcatalogue} will be of service to the community. In particular, our Appendix \ref{GRBcatalogue} should represent a tangible improvement over both~\cite{icecubeGRBs} and~\cite{maxplanckGRBs}  for researchers exploring alternative search strategies for GRB neutrinos affected by in-vacuo dispersion.

\section{Simulated data for the statistical-significance analysis and background estimate}
\label{sec:background}

As stressed above, the in-vacuo-dispersion significance of our findings can only be investigated using a statistical approach based on the $p$-value, {\it i.e.} estimating numerically the  probability that one could find GRB-neutrino candidates in such good agreement with the in-vacuo-dispersion hypothesis just by accident, as an outlier of the type of findings one would expect within the null hypothesis $\eta = 0$ of no in-vacuo dispersion.

As done in the previous neutrino studies\footnote{The interested reader can find applications of this method of statistical analysis to studies not involving neutrinos, {\it e.g.}, in Refs.~\cite{LIGOScientific:2017zic, reshufflingFERMI, reshufflingLIV, reshufflingMAGIC, reshufflingFiorentini, reshufflingMed}.} of Refs.~\cite{Ryan,RyanLensing,NatureAstro}, we do this by producing simulated data through transformations of the real data which reliably preserve their morphology while washing away any possible correlations of the type predicted by Eq.~(\ref{mainnewone}) between the time of arrival, the neutrino energy, and $D(z)$.

Our simulated data are obtained from the real data performing the following independent manipulations:

\noindent
$\bullet$ we act on the neutrino observation times with a random permutation and a random periodic time translation (the periodicity makes it sure that they stay within the actual IceCube observation window);

\noindent
$\bullet$ we act on the neutrino directions with a random rotation around the Earth's axis;

\noindent
$\bullet$ we act on the GRB directions with a random permutation and a random rotation around the Galactic axis.

A combination of these manipulations is arguably the most general transformation of the true data not affecting their morphology in any relevant respect. In fact:

\noindent
$\bullet$ the neutrino observation times are statistically compatible with their being uniformly distributed within the IceCube observation window, and there are no a-priori reason for expecting a change in the HESE neutrino rate in this period;

\noindent
$\bullet$ the efficiency of the IceCube detector, whose axis is closely aligned to the Earth's axis, is virtually independent of the right ascension, but not of the declination, of incoming neutrinos, as clearly evidenced in the neutrino angular data; we choose not to reshuffle the neutrino directions so as not to spoil their evident correlation with the neutrino energies (the higher the neutrino energy, the more accurate its reconstructed direction);

\noindent
$\bullet$ the directions of known-redshift GRBs are independent of galactic longitude, but are correlated to some extent with galactic latitude, as the dust lying around the galactic plane makes it much more difficult to determine the redshift of a GRB in that region; also, we do not act on the GRB observation times because we find that the detection rate of known-redshift GRBs is  declining with time, probably another selection effect due to older events having been more likely followed by the long-term, host-galaxy observations which are often needed to determine the redshift of a GRB.

As a first application of our simulated data we can estimate the overall background that is expected with the criteria adopted by our strategy of analysis. For this purpose, we generated $10^5$ instances of simulated data\footnote{In the following, whenever we refer to simulated data, it is understood that we did all computations based on $10^5$ instances.} and we used them to estimate the expected number of HESE shower neutrinos that in the absence of in-vacuo dispersion could accidentally be associated with a GRB of known redshift according to the time-energy and direction compatibility conditions specified above (with $\eta\in[12.7,30.7]$ and $\mathcal{S}_{cut}=1/4\pi$). We find that, on average, our criteria should pick up 1.02 such accidental GRB-neutrino associations.

\section{Main analysis}\label{main}
\begin{figure}[H]
    \centering
\includegraphics[scale=0.46]{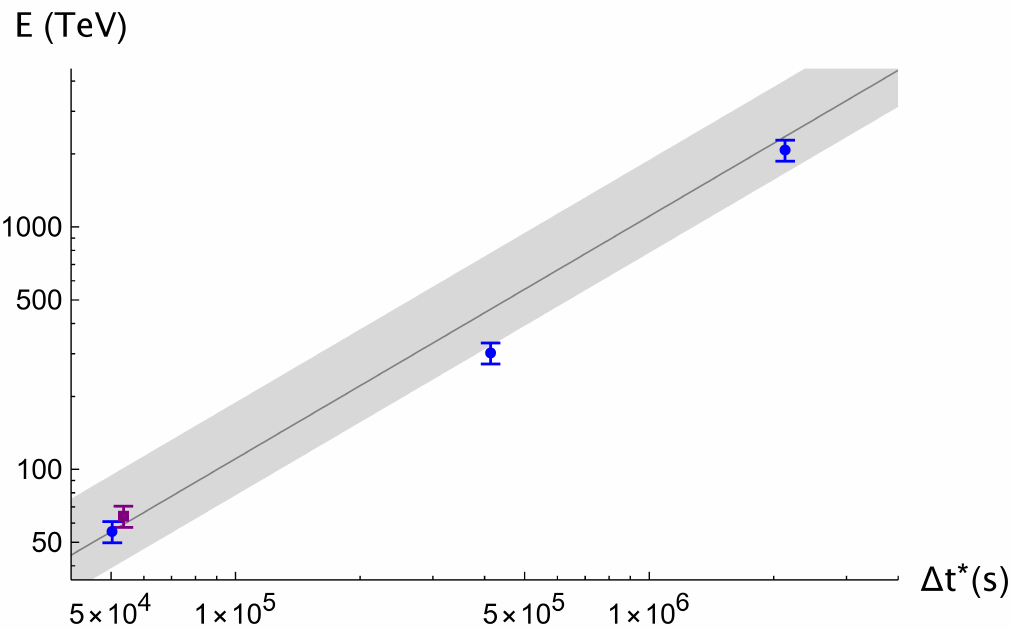}
\caption{The values of $E$ and $\Delta t^*$ of our four GRB-neutrino candidates line up rather nicely according to the expectations of in-vacuo dispersion. Only one of them, here shown as a violet square, was taken into account also in the analysis of Ref.~\cite{natastro2023}.
The content of this figure should be assessed also keeping in mind the results here reported in Sec. III, suggesting that it is likely that  one of our four GRB-neutrino candidates is background.
Based on the estimate $\eta = 21.7 \pm 4.5$ reported in Ref.~\cite{natastro2023}, the dark-gray line corresponds to $\eta = 21.7$, while the light-gray band highlights our region of interest in the $E$-$\Delta t^*$ plane, corresponding to $\eta\in[12.7,30.7]$ via Eq.~(\ref{mainnewone}) (see Sec.~\ref{sec:intro} for comments on this interval). }
\label{fig:1}
\end{figure}

Equipped with the preparatory work reported in the previous sections, we are now ready to discuss our main analysis.
We begin by looking for  GRB-neutrino candidates, using the data described in Sec.~\ref{sec:data}. 
For each GRB-neutrino pair selected by our directional criteria ($\mathcal{S} \geq 1/4\pi$)
we describe the  $\Delta t$ of Eq.~(\ref{mainnewone})
as the difference between the neutrino observation time\footnote{If there was no in-vacuo dispersion, a GRB neutrino would
be observed (nearly-)simultaneously with the GRB photons. We attribute the whole of the time-of-arrival difference 
to the $\Delta t$ of the neutrino, since photons observed from GRBs are of much lower energies than our neutrinos
and the effect we are studying depends linearly on energy. Some quantum-spacetime scenarios predict in-vacuo dispersion with different magnitude for photons and neutrinos (and if the magnitude was much stronger for photons this assumption of our analysis would not be satisfied); it is however noteworthy that effective-field-theory approaches to quantum properties of spacetime predict that the type of pure in-vacuo dispersion here studied is actually forbidden for photons whereas it is allowed for neutrinos~\cite{Mattingly:2005re, Stecker:2014oxa}.}
and the observation time of the GRB, which we label as
$\Delta t_{candidate}$, and then we use Eq.~(\ref{mainnewone}) and the known redshift of the GRB, which we label as $z_{GRB}$, to convert the requirement that $\eta$
should be within the interval 
$[12.7,30.7]$ into a corresponding range 
of allowed values for the neutrino energy. We then  consider the GRB-neutrino pair a GRB-neutrino candidate if the energy of the neutrino is compatible with that energy range within two standard deviations
(assuming 10\% uncertainty in the energy of the neutrino~\cite{IceCubeEnergy2013}).
We find that there are four such GRB-neutrino candidates.
Fig.~\ref{fig:1} provides a visual characterization of the fact that the properties of these four GRB-neutrino candidates 
match rather well the expectations of the in-vacuo-dispersion scenario of Eq.~(\ref{mainnewone}). 

In Fig.~\ref{fig:1},  $\Delta t^*$ is defined as $$\Delta t^* \equiv {D(1) \over D(z_{GRB})} \Delta t_{candidate}$$
(the numerical factor  $D(1)$ is introduced only for the convenience of having a  $\Delta t^*$ with dimensions of time), and
according to
Eq.~(\ref{mainnewone}) one should find a linear relationship between $E$
and $\Delta t^*$, 
$$\Delta t^*= \eta D(1) \frac{E}{M_P} \, .$$
It is clearly encouraging for the hypothesis of in-vacuo dispersion that our criteria find four GRB-neutrino candidates (some details on these four GRB-neutrino candidates are provided in Appendix \ref{sec:GRBneutprop}), whereas  according to the pure-background null hypothesis we expected to typically find only one (see Sec.~\ref{sec:background}). Further encouragement comes from the fact that in Fig.~\ref{fig:1} our four GRB-neutrino candidates line up rather nicely, as measured by their correlation coefficient $r(\Delta t^*,E) =0.9985 $. In particular, three of the four GRB-neutrino candidates are compatible with a remarkably narrow range of values of $\eta$, an observation which strengthens the appeal of the conjecture that three of our four GRB-neutrino candidates actually might be GRB neutrinos affected by in-vacuo dispersion. 

The fact that our findings reported in Fig.~\ref{fig:1} fit very well with the estimate $\eta = 21.7 \pm 4.5$, 
which resulted from the analysis of Ref.~\cite{natastro2023}, is noteworthy since, because of the different criteria adopted for the selection of GRB-neutrino candidates, the two analyses only share one GRB-neutrino candidate, the one shown as a  violet square in Fig.~\ref{fig:1}.
The analysis leading to the estimate 
$\eta = 21.7 \pm 4.5$, 
in Ref.~\cite{natastro2023} ended up focusing on seven GRB-neutrino candidates, but only one of them involved a GRB with measured redshift.

It is of course necessary to quantify statistically this observation that the data we are analyzing match well the expectations of in-vacuo dispersion. For this purpose we rely on the same strategy of characterization of statistical significance adopted in Ref.~\cite{natastro2023}.
In parts of the next Section we shall contemplate alternative strategies of characterization of statistical significance, but the main objective of the study we are here reporting is to compare the GRB-neutrino-candidate selection criteria here adopted with the selection criteria  adopted in Ref.~\cite{natastro2023}, and this comparison is of course more transparent if the two selection strategies are assessed using the same characterization of statistical significance.
The main analysis of 
Ref.~\cite{natastro2023}
focused on seven GRB-neutrino candidates 
and  established 
how frequently simulated data produced at least seven GRB-neutrino candidates with correlation at least as high as the correlation of the GRB-neutrino candidates found in the true data. This happened in about 0.7\% of the trials, yielding the above-mentioned $p$-value  $P_{[34]}=0.007$.
Proceeding in an analogous way with the selection criteria here adopted, we start by using our simulated data (Sec. III) to estimate how frequently the null hypothesis of no in-vacuo dispersion ($\eta =0$) would produce at least four GRB-neutrino candidates according to our new selection criteria, finding a $p$-value $P_{n}=0.019$. 
We then proceed to estimate how frequently the null hypothesis would accidentally exhibit at least four GRB-neutrino candidates with correlation $r(\Delta t^*,E)\geq0.9985$, finding a $p$-value $P_{r}=0.006$ (which corresponds to a $2.8\sigma$ significance in Gaussian statistics).

\section{Avenues for refining the strategy of analysis}\label{sec:refinements}
We consider as our main result
the $p$-value $P_{r}=0.006$ derived in the previous Section, which, while  being still far from conclusive, testifies to the discovery potential of the novel GRB-neutrino-candidate selection criteria here advocated, which was our main objective.

In this Section we discuss and investigate some observations which could be used  to refine the strategy of analysis. We believe that some of the observations reported in this Section deserve being considered in setting up in-vacuo-dispersion analyses of future neutrino data, even though some of the quantitative aspects of this Section should be assessed with caution since they were produced in a second phase of our investigations after seeing the data, and might therefore be affected by the sources of bias that are well known for unblind analyses.   
In particular, some of the observations reported in this Section are at least in part inspired by the fact that we saw that among our four GRB-neutrino candidates there are three that are described particularly well 
according to Eq.~(\ref{mainnewone}), and by our perception 
that the procedure used in Sec. IV to estimate the significance of our findings did not ``benefit" sufficiently from the noteworthy properties of those three best GRB-neutrino candidates.

\subsection{Alternative choices of $S_{cut}$}
In planning our study we ended up adopting the rather prudent criterion of directional consistency characterized by the intuitively natural choice $\mathcal{S}_{cut} = 1/4\pi$ (see Sec.~\ref{sec:intro}). After completing our main analysis we perceived the need to quantify how frequently this criterion was picking up accidental directional associations between a neutrino and a GRB. Using our simulated data, we checked that the choice $\mathcal{S}_{cut}= 1/4\pi$ entails a 7.8\% probability of accidental directional association between a neutrino and a GRB, which is of course partly responsible for the (large but) tolerable expectation that about one background GRB-neutrino candidate should be accidentally picked up by our overall selection criteria (see Sec. III).

\begin{figure}[h!]
    \centering
\includegraphics[scale=0.65]{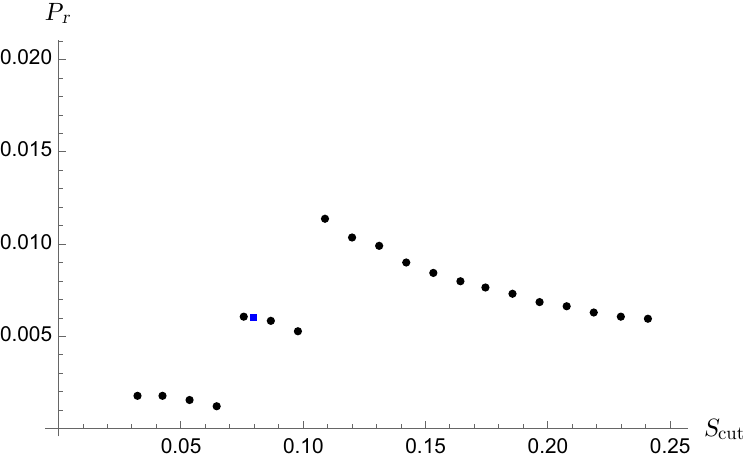}
\caption{The $p$-value $P_{r}$ (see Sec.~\ref{main} for the definition) as a function of $\mathcal{S}_{cut}$. The blue square corresponds to the choice $\mathcal{S}_{cut}=1/4\pi$ of Sec.~\ref{main}.}
    \label{fig:Scut}
\end{figure}

We then checked how strongly the $p$-value $P_{r}$, which we regard as our main result, depends on the choice of $\mathcal{S}_{cut}$. Fig.~\ref{fig:Scut} shows the dependence of $P_{r}$ on $\mathcal{S}_{cut}$ for $\mathcal{S}_{cut}\in[0.032,0.241]$. The rationale for exploring this particular range of values is that, as checked using our simulated data, 
$\mathcal{S}_{cut}=0.241$ entails a 4.5\% probability of accidental directional association, the nominal ``two standard deviation" reference, whereas $\mathcal{S}_{cut}=0.032$ corresponds to a 10\% probability of spurious association.
As clearly evidenced by Fig.~\ref{fig:Scut}, our main result $P_{r}$ does not depend strongly on the choice of $\mathcal{S}_{cut}$. It is potentially intriguing that for values of $\mathcal{S}_{cut}$ just below $1/4\pi$ we would have found an even smaller $P_{r}$,
but, in keeping with the considerations of Sec.~\ref{sec:intro}, we currently regard this circumstance as a mere numerical accident occurring with the presently-available data (rather than a hint for optimizing the analysis of future data).

\subsection{Improved correlation}
One challenge whose handling might require further methodological refinements in future studies concerns the choice of an appropriate statistic quantifying the agreement between the properties of the found GRB-neutrino candidates and Eq.~(\ref{mainnewone}). Investigations of this aspect should take into account the fact that the time window for in-vacuo dispersion searches of GRB neutrinos will always be rather large, even if, as one might hope, the $\eta$ range on which one would focus could become narrower with the accrual of more data. A relatively small uncertainty on $\eta$ still translates (especially at high neutrino energies and high redshift) into a rather large time window, large enough for a tangible probability of background contamination of the GRB-neutrino candidates. 
These concerns are even more severe if one allows for non-systematic, fuzzy in-vacuo dispersion effects of the type mentioned here briefly in Sec.~\ref{sec:intro} (and discussed in more detail in, {\it e.g.}, Refs.~\cite{GAClivingReview, COSTreview, gacPIRANnaturephysics2015}), since in that case searches of GRB-neutrino candidates would always have to explore a range of $\eta$ with width at least $\delta \eta$.

Awareness of these issues should encourage investigations aimed at quantifying the agreement between the found GRB-neutrino candidates and the relevant in-vacuo-dispersion models in ways that are both robust against contamination by the background and capable of profiting from all the implications of in-vacuo-dispersion.

In our main analysis, reported in Sec.~\ref{main}, we followed Ref.~\cite{natastro2023} and characterized 
the agreement between our findings and Eq.~(\ref{mainnewone}) in terms of the standard (Pearson) correlation coefficient
$r(\Delta t^*,E)$. However,  $r$, while giving an appropriate weight to the linear relationship between $E$ and $\Delta t^*$ predicted by Eq.~(\ref{mainnewone}), is completely insensitive to the fact that, according to Eq.~(\ref{mainnewone}), one should have $\Delta t^* \rightarrow 0$ for $E \rightarrow 0$. We observe that this shortcoming can be remedied by replacing the sample covariance and the sample variances appearing in the definition of $r$ with the corresponding raw (non-central) sample moments, \emph{i.e.}, by replacing $r(\Delta t^*,E)$ with the corresponding non-central correlation coefficient 
$$r_{0}(\Delta t^*,E)=\frac{\sum_i\Delta t^*_i E_i}{\sqrt{\sum_j(\Delta t^*_j)^2\sum_kE_k^2}}.$$
It is clear that $r_{0}$ indeed rewards GRB-neutrino candidates fitting the expectation that $E$ depends linearly on $\Delta t^*$ with negligible intercept. 

We can test the effectiveness of this improved statistic by replacing $r$ with $r_{0}$ in our main analysis (Sec.~\ref{main}). Our four GRB-neutrino candidates have $r_{0}(\Delta t^*,E)=0.9989$, and requiring that our simulated data (Sec.~\ref{sec:background}) accidentally produce at least four GRB-neutrino candidates with $r_{0}(\Delta t^*,E)\geq0.9989$, we find a $p$-value $P_{r_0}=0.004$.

The fact that $P_{r_0}<P_{r}$ suggests that our GRB-neutrino candidates are indeed consistent with the expectation that $\Delta t^* \rightarrow 0$ for $E \rightarrow 0$. 

\subsection{Taking into account the background estimate}
Regardless of the statistic used to quantify the agreement of the found GRB-neutrino candidates with Eq.~(\ref{mainnewone}), one could also try to tame background contamination by taking explicitly into account its contribution before computing the statistic.

Looking at the GRB-neutrino candidate with $E=302$~TeV in Fig.~\ref{fig:1}, one can easily understand these concerns: even though we expect that one of our four GRB-neutrino candidates should be background, in our main analysis (following Ref.~\cite{natastro2023}) we computed the correlation coefficient of all of them. Future studies might thus contemplate the possibility of taking into account the background estimate. More explicitly, if one finds $N$ GRB-neutrino candidates and estimates that there is a large probability, {\it e.g.}, 90\%,  that at least $M$ should be background, it may be appropriate to assess the agreement between the found GRB-neutrino candidates and Eq.~\eqref{mainnewone} using only the best $N-M$ candidates.

In the case of the study here reported, the probability of finding at least one background candidate is still too low (about 60\%) to justify disregarding our worst candidate. Nevertheless, for mere illustrative purposes, we briefly discuss how to amend our main analysis if one of our four candidates could be actually ignored. First, we compute the correlation coefficients of all possible choices of three of our four GRB-neutrino candidates and find that the highest correlation is $r(\Delta t^*,E)=0.999997$ (found indeed excluding the candidate with $E=302$~TeV).
We then use our simulated data to estimate how frequently the null hypothesis would accidentally produce at least four GRB-neutrino candidates such that three of them have correlation $r(\Delta t^*,E)\geq0.999997$, finding a $p$-value $P_{rb}=0.002$.

\subsection{Scanning a range of values of $\eta$}

The fact that we are restricting our analysis to GRBs with measured redshift allows us to explore another statistic, which we expect to be less vulnerable than correlation to background contamination.

\begin{figure}[H]
\centering
\includegraphics[scale=0.55]{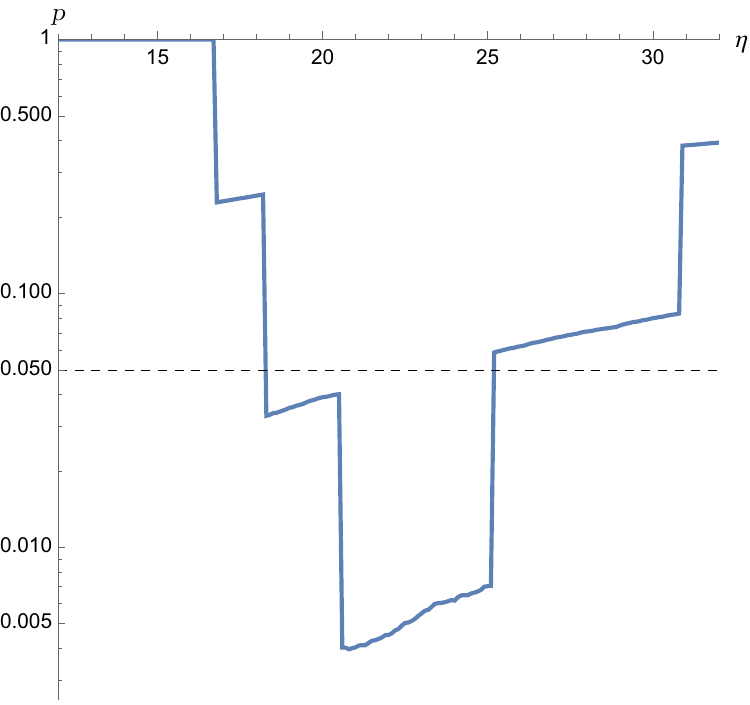}
\caption{The $p$-curve for $\eta\in[12.7,30.7]$. It is obtained estimating how frequently simulated data produce at least the same number of GRB-neutrino candidates as the real data for each value of $\eta$ (see text for details).
The horizontal dashed line is drawn at $p=0.05$.}
\label{fig:pValue}
\end{figure}

For simplicity we illustrate and discuss the strategy of analysis based on this alternative statistic applying it directly to the data already analyzed in this paper.
For every given value of $\eta$ in our interval of interest $[12.7,30.7]$, we find the number $N(\eta)$ of GRB-neutrino candidates which are directionally compatible in the usual sense ($\mathcal{S}\geq1/4\pi$) and satisfy Eq.~\eqref{mainnewone} for that specific value of $\eta$ within two standard uncertainties in the neutrino energy. We then use simulated data to estimate the probability that the null hypothesis would accidentally produce at least $N(\eta)$ such GRB-neutrino candidates, finding a corresponding (local) $p$-value $p(\eta)$. The resulting ``$p$-curve”
is reproduced here as Fig.~\ref{fig:pValue}.

Consistently with the fact, already visible in Fig.~\ref{fig:1}, that our four GRB-neutrino candidates are remarkably compatible with a narrow range of values of $\eta$, we see that the $p$-curve of Fig.~\ref{fig:pValue} shows a deep valley surrounded by considerably higher values of $p$. In particular, $p(\eta)$ is less than 0.05 (marked as a dashed horizontal line in Fig.~\ref{fig:pValue}) for $\eta\in[18.3,25.1]$. The minimum of the $p$-curve is found at $\eta={20.8}$, fully consistent with the estimate $21.7\pm 4.5$ reported in Ref.~\cite{natastro2023}.

The statistical significance of our $p$-curve can be characterized estimating how frequently the minimum of the analogous $p$-curves built out of simulated data happens to be less than or equal to the minimum value $p(20.8)=0.004$ attained in Fig.~\ref{fig:pValue}. The resulting (global) $p$-value is $P_{ns}=0.012$.

The fact that $P_{ns}$ and $P_{r}$, resulting from considerably different strategies of analysis, are of the same order of magnitude suggests that both values are in fact genuine characterizations of the statistical significance of our results, \emph{i.e.}, that they are not too dependent on the specific properties of the corresponding statistics.

\subsection{Exploring $\eta$ outside the range $[12.7,30.7]$}

\begin{figure}[H]
    \centering
\includegraphics[scale=0.59]{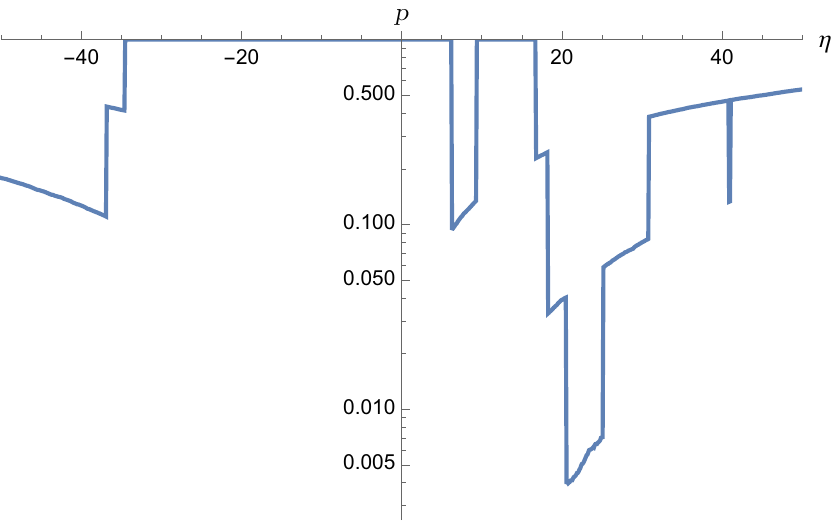}
\caption{The same $p$-curve of Fig.~\ref{fig:pValue} for $\eta\in[-50,50]$.}
    \label{fig:pValue2}
\end{figure}

In our main analysis we let $\eta$ vary in the range $[12.7,30.7]$ favored by the previous study \cite{natastro2023}, which adopted different selection criteria, obtaining results in remarkable agreement with those of Ref.~\cite{natastro2023}. Still, it is interesting to check whether our new GRB-neutrino-candidate selection criteria would have yielded even stronger results for values of $\eta$ outside the range $[12.7,30.7]$.
A natural way of doing this is to extend the domain of the $p$-curve introduced in the previous Subsection. In Fig. \ref{fig:pValue2} we show the $p$-curve for $\eta\in[-50,50]$.

Looking at Fig.~\ref{fig:pValue2}, it is clear that also our new GRB-neutrino-candidate selection criteria favor the range 
$[12.7,30.7]$ for $\eta$.
The fact that we find no other values of $\eta$ with comparable $p(\eta)$ is also potentially significant
for a much-studied effective-field-theory description of quantum-spacetime properties~\cite{Colladay:1998fq, Stecker:2014oxa},  which in principle allows independent in-vacuo-dispersion parameters for the two neutrino helicities. The prevalent theoretical prejudice is that the two helicities should have the same in-vacuo-dispersion parameter, but it has also been argued that the two helicities might have $\eta$ with the same modulus but opposite sign~\cite{Ryan, MaNeutriniFirst}. The $p$-curve reproduced in Fig.~\ref{fig:pValue2} clearly does not provide any encouragement for this helicity-dependence scenario.

\section{Outlook}

The essential equivalence of the $p$-values $P_{[34]}$, characterizing the statistical significance of the previous analysis~\cite{natastro2023}, and $P_{r}$, found here in Sec.~\ref{main} and characterizing in the same way the significance of our current analysis, shows that the new, redshift-based selection criteria for GRB-neutrino candidates proposed in Sec.~\ref{sec:intro} are competitive with the ones adopted in Ref.~\cite{natastro2023}. This means that the improvement in data quality provided by the knowledge of the GRB redshifts is more than sufficient to compensate for the loss of most of the GRB dataset (we measure the redshift of about 1/10 of observed GRBs). We feel that this is our main result, as it effectively opens up a new avenue, complementary to the approach pursued in Ref.~\cite{natastro2023}, to investigate in-vacuo dispersion for neutrinos.

Considering the very significant differences in their selection criteria (as stressed above, only one GRB-neutrino candidate figures in both analyses), it is noteworthy that the analysis here reported and the one reported in Ref.~\cite{natastro2023} both favor the same range of values of $\eta$. This might suggest that, if one were somehow able to combine the two analyses, then the overall evidence in favor of in-vacuo dispersion provided by presently-available IceCube data might be stronger than indicated by the individual statistical significance of each analysis. However, we do not yet see how such a combination of selection criteria could be achieved in a logically consistent manner, mostly because sharp knowledge of the GRB redshifts is a structural feature of the strategy of analysis advocated here.

We expect that the redshift-based strategy articulated in the previous sections will prove to be the best starting point for any future refinements, but we do not exclude that a suitable way may still be found for including GRBs whose redshift has not been measured (of course, from the perspective of our approach, somehow estimating the redshift of those GRBs with a suitably large uncertainty). Such a refinement of our approach could prove even more valuable in a few years, as more IceCube and KM3NeT data is accrued.

At least equally important for the analysis of future IceCube and KM3NeT data would be to remove the other key limitation of our selection criteria, the restriction to shower-event neutrinos.
This could in principle be achieved by incorporating into the analysis the known~\cite{jackJrEnergy2018} large and non-Gaussian probability distribution (whose peak lies sizably above the visible energy) that relates the primary neutrino energy to the visible energy in track events.

\begin{acknowledgments}
We thank Anastasia Tsvetkova for guidance on the available GRB catalogues and resources, which was very valuable in the early stages of our work on Appendix~\ref{GRBcatalogue}. G.A.-C. and G.G. are grateful for financial support by the Programme STAR Plus, funded by Federico II University and Compagnia di San Paolo,
 and by the MIUR, PRIN 2017
grant 20179ZF5KS. G.R.’s work on this project was
supported by the National Science Centre grant
2019/33/B/ST2/00050. G.D.’s work on this project was supported by the Research Council of Norway, project number 301718, and by the Beatriu de Pin\'{o}s programme (BP 2023). G.F.'s work on this project was supported by ``Fondazione Angelo Della Riccia" and ``The Foundation Blanceflor".
This work also falls within the
scopes of the  COST Action CA23130 “Bridging high and low energies in search of quantum gravity” and the COST Action CA18108 "Quantum Gravity phenomenology in the Multi-Messenger era".
\end{acknowledgments}

\appendix 

\section{Known-redshift GRB data}\label{GRBcatalogue}

The following table contains the GRB data used in our analysis. We include all GRBs observed up to December 2022 for which a definite redshift determination is available in the literature. We did not include GRBs with only upper and/or lower bounds on redshift. For the selection of the sample and the redshift values $z$ we relied mostly on Greiner's catalogue~\cite{maxplanckGRBs}, referred to in our list as MP. Searching the literature for other  GRBs whose redshift was measured, we found 45 which are not acknowledged as such in the summary table of~\cite{maxplanckGRBs}. For these GRBs we provide references to the relevant works and GCNs in the column ``$z$ ref." of our table.

The angular data and the trigger times (TT) of the GRBs in our sample were downloaded from the web catalogue maintained by the IceCube collaboration~\cite{icecubeGRBs}, and were then manually checked against the primary GCN sources referred to in Greiner's list~\cite{maxplanckGRBs}. If the right ascension (RA) or the declination (decl) values reported by \cite{icecubeGRBs} were no more than 0.001° off the best values we found in~\cite{maxplanckGRBs}, we did not correct them, as our analysis is completely insensitive to errors of that order of magnitude. For the same reason, we do not provide the uncertainties affecting the angular positions, since we checked that they are all less than 0.001°. Accordingly, we report our angular values with three digital figures, with the understanding that the last one is affected by an uncertainty of order 0.001°.

\setlength{\tabcolsep}{3.2pt}
\begin{longtable}{@{}lrrcll@{}}
\caption{List of GRBs with measured redshift}
\label{TabA} \\
\hline\hline
\rule{0pt}{10pt}\hspace{16pt}Name & RA (°) & decl (°)  & TT (MJD) & \hspace{8pt}$z$ &  $z$ ref. \\
\hline\hline
\endfirsthead

\hline
\rule{0pt}{10pt}\hspace{16pt}Name & RA (°) & decl (°)  & TT (MJD) & \hspace{8pt}$z$ &  $z$ ref. \\
\hline\vspace{0pt}
\endhead

\hline
\endfoot

&&&&&\\
GRB221226B	&	 22.909	&	-41.527	&	59939.945	&	2.694	&	MP	\\
GRB221110A	&	 29.100	&	-27.294	&	59893.103	&	4.06	&	MP	\\
GRB221009A	&	288.264	&	 19.773	&	59861.553	&	0.151	&	MP	\\
GRB221006A	&	337.246	&	 15.714	&	59858.040	&	0.731	&	\cite{2022GCN.32710....1O,2022GCN.32742....1C}	\\
GRB220813A	&	 81.532	&	-33.016	&	59804.808	&	0.82	&	\cite{2022GCN.32471....1F}	\\
GRB220627A	&	201.369	&	-32.426	&	59757.890	&	3.084	&	MP	\\
GRB220611A	&	 66.515	&	-37.260	&	59741.751	&	2.3608	&	MP	\\
GRB220527A	&	323.528	&	-14.972	&	59726.387	&	0.857	&	MP	\\
GRB220521A	&	275.230	&	 10.372	&	59720.972	&	5.6	    &	MP	\\
GRB220219B	&	240.914	&	 31.234	&	59629.394	&	0.293	&	MP	\\
GRB220117A	&	 91.572	&	-28.437	&	59596.680	&	4.961	&	MP	\\
GRB220107A	&	169.807	&	 34.171	&	59586.615	&	1.246	&	MP	\\
GRB220101A	&	  1.353	&	 31.769	&	59580.215	&	4.618	&	MP	\\
GRB211227A	&	132.149	&	 -2.735	&	59575.981	&	0.228	&	\cite{2021GCN.31324....1M,GRB211227A}	\\
GRB211211A	&	212.292	&	 27.889	&	59559.549	&	0.073	&	\cite{2021GCN.31235....1L,GRB211211A}	\\
GRB211207A	&	149.624	&	-24.359	&	59555.870	&	2.272	&	MP	\\
GRB211024B	&	154.713	&	 24.568	&	59511.931	&	1.1137	&	MP	\\
GRB211023B	&	170.310	&	 39.136	&	59510.879	&	0.862	&	MP	\\
GRB211023A	&	 73.154	&	 85.324	&	59510.546	&	0.39	&	MP	\\
GRB210919A	&	 80.254	&	  1.312	&	59476.020	&	0.2411	&	MP	\\
GRB210905A	&	309.048	&	-44.440	&	59462.009	&	6.318	&	MP	\\
GRB210822A	&	304.438	&	  5.283	&	59448.388	&	1.736	&	MP	\\
GRB210731A	&	300.305	&	-28.061	&	59426.931	&	1.2525	&	MP	\\
GRB210722A	&	 27.030	&	 -6.347	&	59417.871	&	1.145	&	MP	\\
GRB210702A	&	168.578	&	-36.747	&	59397.797	&	1.16	&	MP	\\
GRB210619B	&	319.718	&	 33.850	&	59385.000	&	1.937	&	MP	\\
GRB210610B	&	243.918	&	 14.399	&	59375.827	&	1.13	&	MP	\\
GRB210610A	&	204.282	&	 14.465	&	59375.628	&	3.54	&	MP	\\
GRB210517A	&	358.224	&	-39.102	&	59351.228	&	2.486	&	MP	\\
GRB210504A	&	222.392	&	-30.534	&	59338.580	&	2.077	&	MP	\\
GRB210420B	&	254.325	&	 42.570	&	59324.774	&	1.4	    &	MP	\\
GRB210411C	&	296.612	&	-39.398	&	59315.629	&	2.826	&	MP	\\
GRB210323A	&	317.947	&	 25.369	&	59296.918	&	0.733	&	MP	\\
GRB210321A	&	 87.895	&	 70.130	&	59294.301	&	1.487	&	MP	\\
GRB210312B	&	155.814	&	 76.869	&	59285.870	&	1.069	&	MP	\\
GRB210222B	&	154.606	&	-14.932	&	59267.943	&	2.198	&	MP	\\
GRB210210A	&	262.771	&	 14.663	&	59255.084	&	0.715	&	MP	\\
GRB210204A	&	117.081	&	 11.410	&	59249.270	&	0.876	&	MP	\\
GRB210116A	&	123.814	&	 -5.867	&	59230.246	&	2.514	&	MP	\\
GRB210112A	&	219.006	&	 33.054	&	59226.067	&	2	    &	\cite{2021GCN.29296....1K,GRB210112A}	\\
GRB210104A	&	103.772	&	 64.676	&	59218.477	&	0.46	&	MP	\\
GRB201221D	&	171.059	&	 42.144	&	59204.963	&	1.045	&	MP	\\
GRB201221A	&	214.480	&	-45.416	&	59204.298	&	5.7	    &	MP	\\
GRB201216C	&	 16.370	&	 16.516	&	59199.963	&	1.1	    &	MP	\\
GRB201104B	&	  5.215	&	  7.842	&	59157.732	&	1.954	&	MP	\\
GRB201103B	&	 42.185	&	 12.137	&	59156.755	&	1.105	&	MP	\\
GRB201024A	&	125.952	&	  3.354	&	59146.117	&	0.999	&	MP	\\
GRB201021C	&	 12.529	&	-55.866	&	59143.852	&	1.07	&	MP	\\
GRB201020B	&	 75.470	&	 77.068	&	59142.732	&	0.804	&	MP	\\
GRB201020A	&	261.228	&	 31.428	&	59142.241	&	2.903	&	MP	\\
GRB201015A	&	354.319	&	 53.416	&	59137.952	&	0.426	&	MP	\\
GRB201014A	&	 20.796	&	 27.660	&	59136.950	&	4.56	&	MP	\\
GRB200829A	&	251.205	&	 72.329	&	59090.582	&	1.25	&	MP	\\
GRB200826A	&	  6.786	&	 34.027	&	59087.187	&	0.7481	&	MP	\\
GRB200613A	&	153.042	&	 45.754	&	59013.229	&	1.227	&	MP	\\
GRB200524A	&	213.043	&	 60.905	&	58993.211	&	1.256	&	MP	\\
GRB200522A	&	  5.682	&	 -0.283	&	58991.487	&	0.554	&	MP	\\
GRB200411A	&	 47.664	&	-52.318	&	58950.187	&	0.7	    &	MP	\\
GRB200219A	&	342.638	&	-59.120	&	58898.317	&	0.48	&	MP	\\
GRB200205B	&	107.788	&	-56.488	&	58884.807	&	1.465	&	MP	\\
GRB191221B	&	154.830	&	-38.158	&	58838.861	&	1.148	&	MP	\\
GRB191031D	&	283.289	&	 47.644	&	58787.891	&	0.5	    &	MP	\\
GRB191019A	&	340.025	&	-17.328	&	58775.634	&	0.248	&	MP	\\
GRB191011A	&	 44.728	&	-27.845	&	58767.192	&	1.722	&	MP	\\
GRB191004B	&	 49.205	&	-39.634	&	58760.898	&	3.503	&	MP	\\
GRB190919B	&	311.877	&	-44.695	&	58745.991	&	3.225	&	MP	\\
GRB190829A	&	 44.544	&	 -8.958	&	58724.830	&	0.0785	&	MP	\\
GRB190719C	&	240.207	&	 13.000	&	58683.624	&	2.469	&	MP	\\
GRB190627A	&	244.828	&	 -5.289	&	58661.471	&	1.942	&	MP	\\
GRB190613A	&	182.529	&	 67.235	&	58647.172	&	2.78	&	MP	\\
GRB190324A	&	 49.616	&	-47.215	&	58566.947	&	1.1715	&	MP	\\
GRB190114C	&	 54.505	&	-26.946	&	58497.873	&	0.425	&	MP	\\
GRB190114A	&	 65.544	&	  2.192	&	58497.133	&	3.3765	&	MP	\\
GRB190106A	&	 29.880	&	 23.846	&	58489.566	&	1.859	&	MP	\\
GRB181201A	&	319.297	&	-12.631	&	58453.110	&	0.45	&	\cite{2018GCN.23488....1I,2018GCN.23499....1S}	\\
GRB181123B	&	184.367	&	 14.598	&	58445.231	&	1.754	&	MP	\\
GRB181110A	&	302.318	&	-36.897	&	58432.364	&	1.505	&	MP	\\
GRB181020A	&	 13.982	&	-47.381	&	58411.792	&	2.938	&	MP	\\
GRB181010A	&	 52.570	&	-23.038	&	58401.247	&	1.39	&	MP	\\
GRB180914B	&	332.356	&	 25.062	&	58375.766	&	1.096	&	MP	\\
GRB180805B	&	 25.782	&	-17.494	&	58335.543	&	0.661	&	MP	\\
GRB180728A	&	253.565	&	-54.044	&	58327.728	&	0.117	&	MP	\\
GRB180727A	&	346.666	&	-63.052	&	58326.594	&	2	    &	MP	\\
GRB180720B	&	  0.529	&	 -2.919	&	58319.598	&	0.654	&	MP	\\
GRB180703A	&	  6.468	&	-67.179	&	58302.876	&	0.6678	&	MP	\\
GRB180624A	&	318.098	&	 -2.338	&	58293.576	&	2.855	&	MP	\\
GRB180620B	&	357.521	&	-57.962	&	58289.660	&	1.1175	&	MP	\\
GRB180618A	&	169.941	&	 73.837	&	58287.030	&	0.544	&	MP	\\
GRB180510B	&	 77.969	&	-62.324	&	58248.844	&	1.305	&	MP	\\
GRB180418A	&	170.122	&	 24.933	&	58226.281	&	1.55	&	MP	\\
GRB180404A	&	 83.548	&	-37.168	&	58212.032	&	1    	&	MP	\\
GRB180329B	&	 82.904	&	-23.690	&	58206.589	&	1.998	&	MP	\\
GRB180325A	&	157.428	&	 24.464	&	58202.078	&	2.248	&	MP	\\
GRB180314A	&	 99.265	&	-24.496	&	58191.030	&	1.445	&	MP	\\
GRB180205A	&	126.820	&	 11.542	&	58154.184	&	1.409	&	MP	\\
GRB180115A	&	 12.039	&	-15.630	&	58133.178	&	2.487	&	MP	\\
GRB171222A	&	148.278	&	 35.627	&	58109.684	&	2.409	&	MP	\\
GRB171205A	&	167.415	&	-12.588	&	58092.306	&	0.0368	&	MP	\\
GRB171020A	&	 39.248	&	 15.204	&	58046.963	&	1.87	&	MP	\\
GRB171010A	&	 66.581	&	-10.463	&	58036.792	&	0.3293	&	MP	\\
GRB170903A	&	254.526	&	 34.979	&	57999.534	&	0.886	&	MP	\\
GRB170817A	&	197.450	&	-23.381	&	57982.529	&	0.0098	&	MP	\\
GRB170728B	&	237.981	&	 70.122	&	57962.961	&	1.27	&	MP	\\
GRB170728A	&	 58.888	&	 12.182	&	57962.287	&	1.49	&	MP	\\
GRB170714A	&	 34.350	&	  1.991	&	57948.518	&	0.793	&	MP	\\
GRB170705A	&	191.704	&	 18.307	&	57939.115	&	2.01	&	MP	\\
GRB170607A	&	  7.366	&	  9.243	&	57911.971	&	0.557	&	MP	\\
GRB170604A	&	342.656	&	-15.412	&	57908.798	&	1.329	&	MP	\\
GRB170531B	&	286.884	&	-16.418	&	57904.918	&	2.366	&	MP	\\
GRB170519A	&	163.427	&	 25.374	&	57892.215	&	0.818	&	MP	\\
GRB170428A	&	330.078	&	 26.916	&	57871.384	&	0.454	&	MP	\\
GRB170405A	&	219.828	&	-25.243	&	57848.777	&	3.51	&	MP	\\
GRB170214A	&	256.341	&	 -1.888	&	57798.649	&	2.53	&	MP	\\
GRB170202A	&	152.514	&	  5.012	&	57786.769	&	3.645	&	MP	\\
GRB170127B	&	 19.977	&	-30.358	&	57780.634	&	2.2 	&	MP	\\
GRB170113A	&	 61.733	&	-71.943	&	57766.420	&	1.968	&	MP	\\
GRB161219B	&	 91.714	&	-26.792	&	57741.784	&	0.1475	&	MP	\\
GRB161129A	&	316.228	&	 32.135	&	57721.300	&	0.645	&	MP	\\
GRB161117A	&	322.052	&	-29.614	&	57709.066	&	1.549	&	MP	\\
GRB161108A	&	180.788	&	 24.868	&	57700.148	&	1.159	&	MP	\\
GRB161104A	&	 77.894	&	-51.460	&	57696.404	&	0.79	&	MP	\\
GRB161023A	&	311.022	&	-47.663	&	57684.944	&	2.708	&	MP	\\
GRB161017A	&	142.769	&	 43.127	&	57678.744	&	2.013	&	MP	\\
GRB161014A	&	332.648	&	  7.469	&	57675.522	&	2.823	&	MP	\\
GRB161001A	&	 71.920	&	-57.261	&	57662.045	&	0.67	&	MP	\\
GRB160821B	&	279.977	&	 62.392	&	57621.937	&	0.16	&	MP	\\
GRB160804A	&	221.630	&	  9.999	&	57604.064	&	0.736	&	MP	\\
GRB160629A	&	  4.863	&	 76.967	&	57568.930	&	3.332	&	MP	\\
GRB160625B	&	308.598	&	  6.919	&	57564.945	&	1.406	&	MP	\\
GRB160624A	&	330.193	&	 29.644	&	57563.477	&	0.483	&	MP	\\
GRB160623A	&	315.298	&	 42.221	&	57562.208	&	0.367	&	MP	\\
GRB160509A	&	311.754	&	 76.108	&	57517.374	&	1.17	&	MP	\\
GRB160425A	&	280.327	&	-54.360	&	57503.977	&	0.555	&	MP	\\
GRB160410A	&	150.685	&	  3.478	&	57488.215	&	1.717	&	MP	\\
GRB160408A	&	122.625	&	 71.128	&	57486.268	&	1.9 	&	MP	\\
GRB160327A	&	146.702	&	 54.013	&	57474.386	&	5   	&	MP	\\
GRB160314A	&	112.790	&	 17.000	&	57461.481	&	0.726	&	MP	\\
GRB160303A	&	168.701	&	 22.742	&	57450.455	&	1   	&	MP	\\
GRB160228A	&	107.316	&	 26.932	&	57446.732	&	1.64	&	MP	\\
GRB160227A	&	194.808	&	 78.679	&	57445.814	&	2.38	&	MP	\\
GRB160203A	&	161.951	&	-24.789	&	57421.092	&	3.52	&	MP	\\
GRB160131A	&	 78.168	&	 -7.050	&	57418.348	&	0.971	&	MP	\\
GRB160121A	&	109.088	&	-23.592	&	57408.577	&	1.96	&	MP	\\
GRB160117B	&	132.195	&	-16.367	&	57404.583	&	0.87	&	MP	\\
GRB151229A	&	329.370	&	-20.732	&	57385.285	&	1.4 	&	MP	\\
GRB151215A	&	 93.584	&	 35.516	&	57371.126	&	2.59	&	MP	\\
GRB151112A	&	  2.054	&	-61.663	&	57338.573	&	4.1 	&	MP	\\
GRB151111A	&	 56.845	&	-44.161	&	57337.356	&	3.5 	&	MP	\\
GRB151031A	&	 83.196	&	-39.122	&	57326.243	&	1.167	&	MP	\\
GRB151029A	&	 38.528	&	-35.386	&	57324.326	&	1.423	&	MP	\\
GRB151027B	&	 76.220	&	 -6.450	&	57322.945	&	4.063	&	MP	\\
GRB151027A	&	272.487	&	 61.353	&	57322.166	&	0.81	&	MP	\\
GRB151021A	&	337.644	&	-33.197	&	57316.062	&	2.33	&	MP	\\
GRB150915A	&	319.658	&	-34.914	&	57280.888	&	1.968	&	MP	\\
GRB150910A	&	  5.667	&	 33.473	&	57275.378	&	1.359	&	MP	\\
GRB150831A	&	221.024	&	-25.635	&	57265.440	&	1.18	&	MP	\\
GRB150821A	&	341.913	&	-57.894	&	57255.406	&	0.755	&	MP	\\
GRB150818A	&	230.356	&	 68.342	&	57252.484	&	0.282	&	MP	\\
GRB150728A	&	292.229	&	 33.916	&	57231.536	&	0.46	&	MP	\\
GRB150727A	&	203.969	&	-18.325	&	57230.793	&	0.313	&	MP	\\
GRB150616A	&	314.717	&	-53.394	&	57189.951	&	1.188	&	\cite{Selsing:2018dwd}	\\
GRB150518A	&	234.201	&	 16.330	&	57160.904	&	0.256	&	MP	\\
GRB150514A	&	 74.876	&	-60.968	&	57156.774	&	0.807	&	MP	\\
GRB150424A	&	152.306	&	-26.631	&	57136.321	&	1   	&	\cite{2015GCN.18100....1T,GRB150424A}	\\
GRB150423A	&	221.579	&	 12.284	&	57135.269	&	1.394	&	MP	\\
GRB150413A	&	190.425	&	 71.841	&	57125.580	&	3.139	&	MP	\\
GRB150403A	&	311.505	&	-62.711	&	57115.913	&	2.06	&	MP	\\
GRB150323A	&	128.178	&	 45.465	&	57104.118	&	0.593	&	MP	\\
GRB150314A	&	126.670	&	 63.834	&	57095.205	&	1.758	&	MP	\\
GRB150301B	&	 89.166	&	-57.970	&	57082.818	&	1.5169	&	MP	\\
GRB150206A	&	 10.074	&	-63.182	&	57059.604	&	2.087	&	MP	\\
GRB150120B	&	 39.291	&	  8.078	&	57042.307	&	3.5 	&	MP	\\
GRB150120A	&	 10.319	&	 33.995	&	57042.123	&	0.46	&	MP	\\
GRB150101B	&	188.020	&	-10.934	&	57023.641	&	0.134	&	MP	\\
GRB141225A	&	138.779	&	 33.792	&	57016.959	&	0.915	&	MP	\\
GRB141221A	&	198.287	&	  8.205	&	57012.338	&	1.452	&	MP	\\
GRB141220A	&	195.066	&	 32.146	&	57011.252	&	1.3195	&	MP	\\
GRB141212A	&	 39.125	&	 18.147	&	57003.510	&	0.596	&	MP	\\
GRB141121A	&	122.669	&	 22.217	&	56982.150	&	1.47	&	MP	\\
GRB141109A	&	144.531	&	 -0.608	&	56970.243	&	2.993	&	MP	\\
GRB141028A	&	322.602	&	 -0.231	&	56958.455	&	2.33	&	MP	\\
GRB141026A	&	 44.084	&	 26.928	&	56956.109	&	3.35	&	MP	\\
GRB141004A	&	 76.734	&	 12.820	&	56934.973	&	0.573	&	MP	\\
GRB140930B	&	  6.348	&	 24.295	&	56930.821	&	1.465	&	MP	\\
GRB140907A	&	 48.146	&	 46.605	&	56907.672	&	1.21	&	MP	\\
GRB140903A	&	238.014	&	 27.603	&	56903.625	&	0.351	&	MP	\\
GRB140808A	&	221.222	&	 49.215	&	56877.037	&	3.29	&	MP	\\
GRB140801A	&	 44.069	&	 30.938	&	56870.792	&	1.32	&	MP	\\
GRB140713A	&	281.106	&	 59.634	&	56851.780	&	0.935	&	MP	\\
GRB140710A	&	 41.068	&	 35.499	&	56848.428	&	0.558	&	MP	\\
GRB140703A	&	 12.996	&	 45.102	&	56841.026	&	3.14	&	MP	\\
GRB140629A	&	248.977	&	 41.877	&	56837.595	&	2.275	&	MP	\\
GRB140623A	&	225.473	&	 81.191	&	56831.223	&	1.92	&	MP	\\
GRB140622A	&	317.173	&	-14.419	&	56830.400	&	0.959	&	MP	\\
GRB140620A	&	281.871	&	 49.731	&	56828.219	&	2.04	&	MP	\\
GRB140614A	&	231.169	&	-79.129	&	56822.045	&	4.233	&	MP	\\
GRB140606B	&	328.125	&	 32.015	&	56814.133	&	0.384	&	MP	\\
GRB140518A	&	227.252	&	 42.418	&	56795.387	&	4.707	&	MP	\\
GRB140515A	&	186.064	&	 15.105	&	56792.384	&	6.32	&	MP	\\
GRB140512A	&	289.370	&	-15.094	&	56789.814	&	0.725	&	MP	\\
GRB140509A	&	 46.594	&	-62.639	&	56786.099	&	2.4 	&	MP	\\
GRB140508A	&	255.466	&	 46.780	&	56785.128	&	1.027	&	MP	\\
GRB140506A	&	276.775	&	-55.636	&	56783.880	&	0.889	&	MP	\\
GRB140430A	&	102.936	&	 23.024	&	56777.857	&	1.6 	&	MP	\\
GRB140428A	&	194.368	&	 28.385	&	56775.945	&	4.7 	&	MP	\\
GRB140423A	&	197.286	&	 49.842	&	56770.355	&	3.26	&	MP	\\
GRB140419A	&	126.990	&	 46.240	&	56766.171	&	3.956	&	MP	\\
GRB140331A	&	134.864	&	  2.717	&	56747.243	&	1      	&	\cite{Chrimes:2018ptj}	\\
GRB140318A	&	184.089	&	 20.209	&	56734.006	&	1.02	&	MP	\\
GRB140311A	&	209.305	&	  0.642	&	56727.879	&	4.954	&	MP	\\
GRB140304A	&	 30.643	&	 33.474	&	56720.557	&	5.283	&	MP	\\
GRB140301A	&	 69.558	&	-34.257	&	56717.642	&	1.416	&	MP	\\
GRB140226A	&	221.492	&	 14.993	&	56714.419	&	1.98	&	MP	\\
GRB140213A	&	105.155	&	-73.137	&	56701.807	&	1.2076	&	MP	\\
GRB140206A	&	145.334	&	 66.761	&	56694.304	&	2.73	&	MP	\\
GRB140129B	&	326.757	&	 26.206	&	56686.536	&	0.43	&	MP	\\
GRB140114A	&	188.522	&	 27.951	&	56671.498	&	3      	&	MP	\\
GRB131231A	&	 10.590	&	 -1.653	&	56657.198	&	0.642	&	MP	\\
GRB131229A	&	 85.232	&	 -4.396	&	56655.277	&	1      	&	MP	\\
GRB131227A	&	 67.378	&	 28.883	&	56653.198	&	5.3	    &	MP	\\
GRB131117A	&	332.331	&	-31.762	&	56613.024	&	4.042	&	MP	\\
GRB131108A	&	156.502	&	  9.662	&	56604.862	&	2.4    	&	MP	\\
GRB131105A	&	 70.967	&	-62.995	&	56601.087	&	1.686	&	MP	\\
GRB131103A	&	348.919	&	-44.640	&	56599.922	&	0.599	&	MP	\\
GRB131030A	&	345.067	&	 -5.368	&	56595.872	&	1.295	&	MP	\\
GRB131011A	&	 32.526	&	 -4.411	&	56576.741	&	1.874	&	MP	\\
GRB131004A	&	296.113	&	 -2.958	&	56569.904	&	0.717	&	MP	\\
GRB130925A	&	 41.179	&	-26.153	&	56560.164	&	0.347	&	MP	\\
GRB130907A	&	215.892	&	 45.608	&	56542.902	&	1.238	&	MP	\\
GRB130831A	&	358.624	&	 29.430	&	56535.545	&	0.4791	&	MP	\\
GRB130822A	&	 27.922	&	 -3.208	&	56526.663	&	0.154	&	MP	\\
GRB130716A	&	179.574	&	 63.053	&	56489.442	&	2.2    	&	MP	\\
GRB130702A	&	217.312	&	 15.774	&	56475.004	&	0.145	&	MP	\\
GRB130701A	&	357.229	&	 36.100	&	56474.179	&	1.155	&	MP	\\
GRB130615A	&	274.829	&	-68.161	&	56458.406	&	2.9    	&	\cite{2013GCN.14898....1E,Selsing:2018dwd}	\\
GRB130612A	&	259.794	&	 16.720	&	56455.141	&	2.006	&	MP	\\
GRB130610A	&	224.420	&	 28.207	&	56453.133	&	2.092	&	MP	\\
GRB130606A	&	249.396	&	 29.796	&	56449.878	&	5.913	&	MP	\\
GRB130604A	&	250.187	&	 68.226	&	56447.288	&	1.06	&	MP	\\
GRB130603B	&	172.201	&	 17.071	&	56446.659	&	0.356	&	MP	\\
GRB130528A	&	139.505	&	 87.301	&	56440.695	&	1.25	&	\cite{Jeong:2014haa}	\\
GRB130518A	&	355.668	&	 47.465	&	56430.580	&	2.489	&	MP	\\
GRB130515A	&	283.440	&	-54.279	&	56427.056	&	0.8    	&	MP	\\
GRB130514A	&	296.283	&	 -7.976	&	56426.301	&	3.6    	&	MP	\\
GRB130511A	&	196.646	&	 18.710	&	56423.480	&	1.3033	&	MP	\\
GRB130505A	&	137.061	&	 17.485	&	56417.349	&	2.27	&	MP	\\
GRB130427B	&	314.898	&	-22.546	&	56409.556	&	2.78	&	MP	\\
GRB130427A	&	173.137	&	 27.699	&	56409.324	&	0.34	&	MP	\\
GRB130420A	&	196.106	&	 59.424	&	56402.311	&	1.297	&	MP	\\
GRB130418A	&	149.037	&	 13.667	&	56400.792	&	1.218	&	MP	\\
GRB130408A	&	134.405	&	-32.361	&	56390.911	&	3.758	&	MP	\\
GRB130215A	&	 43.503	&	 13.395	&	56338.063	&	0.597	&	MP	\\
GRB130131B	&	173.956	&	 15.038	&	56323.799	&	2.539	&	MP	\\
GRB130131A	&	171.127	&	 48.076	&	56323.581	&	1.55	&	MP	\\
GRB121229A	&	190.101	&	-50.594	&	56290.209	&	2.707	&	MP	\\
GRB121226A	&	168.642	&	-30.406	&	56287.798	&	1.37	&	MP	\\
GRB121217A	&	153.710	&	-62.351	&	56278.303	&	3.1    	&	\cite{Elliott:2013tfa}	\\
GRB121211A	&	195.533	&	 30.148	&	56272.574	&	1.023	&	MP	\\
GRB121209A	&	326.787	&	 -8.235	&	56270.916	&	2.1    	&	MP	\\
GRB121201A	&	 13.467	&	-42.943	&	56262.518	&	3.385	&	MP	\\
GRB121128A	&	300.600	&	 54.300	&	56259.212	&	2.2    	&	MP	\\
GRB121123A	&	307.318	&	-11.860	&	56254.419	&	2.7    	&	\cite{2012GCN.13992....1S,2012GCN.14003....1H}	\\
GRB121027A	&	 63.598	&	-58.830	&	56227.314	&	1.773	&	MP	\\
GRB121024A	&	 70.472	&	-12.291	&	56224.122	&	2.298	&	MP	\\
GRB120923A	&	303.795	&	  6.221	&	56193.220	&	7.8    	&	MP	\\
GRB120922A	&	234.748	&	-20.182	&	56192.938	&	3.1    	&	MP	\\
GRB120909A	&	275.736	&	-59.449	&	56179.070	&	3.93	&	MP	\\
GRB120907A	&	 74.750	&	 -9.315	&	56177.017	&	0.97	&	MP	\\
GRB120815A	&	273.958	&	-52.131	&	56154.093	&	2.358	&	MP	\\
GRB120811C	&	199.683	&	 62.301	&	56150.649	&	2.671	&	MP	\\
GRB120805A	&	216.538	&	  5.825	&	56144.895	&	3.1    	&	MP	\\
GRB120804A	&	233.948	&	-28.782	&	56143.038	&	1.05	&	MP	\\
GRB120802A	&	 44.843	&	 13.768	&	56141.334	&	3.796	&	MP	\\
GRB120729A	&	 13.074	&	 49.940	&	56137.456	&	0.8    	&	MP	\\
GRB120724A	&	245.181	&	  3.508	&	56132.277	&	1.48	&	MP	\\
GRB120722A	&	230.497	&	 13.251	&	56130.537	&	0.9586	&	MP	\\
GRB120716A	&	313.051	&	  9.599	&	56124.712	&	2.486	&	MP	\\
GRB120714B	&	355.409	&	-46.184	&	56122.888	&	0.3984	&	MP	\\
GRB120712A	&	169.588	&	-20.034	&	56120.571	&	4.1745	&	MP	\\
GRB120711A	&	 94.678	&	-70.999	&	56119.114	&	1.405	&	MP	\\
GRB120630A	&	352.296	&	 42.556	&	56108.971	&	0.6    	&	MP	\\
GRB120624B	&	170.885	&	  8.929	&	56102.930	&	2.1974	&	MP	\\
GRB120521C	&	214.286	&	 42.145	&	56068.974	&	6      	&	MP	\\
GRB120422A	&	136.910	&	 14.019	&	56039.300	&	0.283	&	MP	\\
GRB120404A	&	235.010	&	 12.885	&	56021.535	&	2.876	&	MP	\\
GRB120401A	&	 58.082	&	-17.636	&	56018.225	&	4.5    	&	\cite{2012GCN.13219....1S}	\\
GRB120327A	&	246.864	&	-29.415	&	56013.122	&	2.813	&	MP	\\
GRB120326A	&	273.905	&	 69.260	&	56012.056	&	1.798	&	MP	\\
GRB120311A	&	273.092	&	 14.296	&	55997.232	&	0.35	&	\cite{Selsing:2018dwd}	\\
GRB120305A	&	 47.536	&	 28.492	&	55991.818	&	0.225	&	MP	\\
GRB120224A	&	 40.942	&	-17.761	&	55981.194	&	1.1    	&	MP	\\
GRB120211A	&	 87.754	&	-24.775	&	55968.499	&	2.4    	&	MP	\\
GRB120119A	&	120.029	&	 -9.082	&	55945.170	&	1.728	&	MP	\\
GRB120118B	&	124.871	&	 -7.185	&	55944.709	&	2.943	&	MP	\\
GRB111229A	&	 76.287	&	-84.711	&	55924.943	&	1.3805	&	MP	\\
GRB111228A	&	150.067	&	 18.298	&	55923.656	&	0.714	&	MP	\\
GRB111225A	&	 13.155	&	 51.572	&	55920.160	&	0.297	&	MP	\\
GRB111215A	&	349.556	&	 32.494	&	55910.586	&	2.06	&	MP	\\
GRB111211A	&	153.090	&	 11.208	&	55906.929	&	0.478	&	MP	\\
GRB111209A	&	 14.344	&	-46.801	&	55904.300	&	0.677	&	MP	\\
GRB111129A	&	307.434	&	-52.713	&	55894.679	&	1.0796	&	MP	\\
GRB111123A	&	154.846	&	-20.645	&	55888.759	&	3.1516	&	MP	\\
GRB111117A	&	 12.693	&	 23.011	&	55882.510	&	2.211	&	MP	\\
GRB111107A	&	129.478	&	-66.520	&	55872.035	&	2.893	&	MP	\\
GRB111008A	&	 60.451	&	-32.709	&	55842.926	&	4.9898	&	MP	\\
GRB111005A	&	223.282	&	-19.737	&	55839.337	&	0.0131	&	MP	\\
GRB110918A	&	 32.539	&	-27.105	&	55822.894	&	0.982	&	MP	\\
GRB110818A	&	317.337	&	-63.981	&	55791.860	&	3.36	&	MP	\\
GRB110808A	&	 57.268	&	-44.194	&	55781.263	&	1.348	&	MP	\\
GRB110801A	&	 89.437	&	 80.956	&	55774.826	&	1.858	&	MP	\\
GRB110731A	&	280.504	&	-28.537	&	55773.465	&	2.83	&	MP	\\
GRB110721A	&	333.659	&	-38.593	&	55763.200	&	0.382	&	\cite{2011GCN.12193....1B,Selsing:2018dwd}	\\
GRB110715A	&	237.684	&	-46.235	&	55757.551	&	0.82	&	MP	\\
GRB110709B	&	164.654	&	-23.455	&	55751.898	&	2.109	&	MP	\\
GRB110503A	&	132.776	&	 52.208	&	55684.733	&	1.613	&	MP	\\
GRB110422A	&	112.046	&	 75.107	&	55673.654	&	1.77	&	MP	\\
GRB110402A	&	197.402	&	 61.253	&	55653.009	&	0.854	&	MP	\\
GRB110213B	&	 41.756	&	  1.146	&	55605.605	&	1.083	&	MP	\\
GRB110213A	&	 42.964	&	 49.273	&	55605.220	&	1.46	&	MP	\\
GRB110205A	&	164.630	&	 67.525	&	55597.085	&	2.22	&	MP	\\
GRB110128A	&	193.896	&	 28.065	&	55589.073	&	2.339	&	MP	\\
GRB110106B	&	134.154	&	 47.003	&	55567.893	&	0.618	&	MP	\\
GRB110106A	&	 79.306	&	 64.174	&	55567.643	&	0.093	&	\cite{2011GCN.11530....1P}	\\
GRB101225A	&	  0.198	&	 44.600	&	55555.776	&	0.847	&	MP	\\
GRB101224A	&	285.924	&	 45.714	&	55554.227	&	0.4536	&	MP	\\
GRB101219B	&	 12.231	&	-34.566	&	55549.686	&	0.5519	&	MP	\\
GRB101219A	&	 74.585	&	 -2.540	&	55549.105	&	0.718	&	MP	\\
GRB101213A	&	241.314	&	 21.897	&	55543.451	&	0.414	&	MP	\\
GRB100906A	&	 28.684	&	 55.630	&	55445.576	&	1.727	&	MP	\\
GRB100905A	&	 31.550	&	 14.930	&	55444.631	&	7.9    	&	MP	\\
GRB100902A	&	 48.629	&	 30.979	&	55441.814	&	4.5    	&	\cite{2010GCN.11195....1C}	\\
GRB100901A	&	 27.264	&	 22.759	&	55440.565	&	1.408	&	MP	\\
GRB100816A	&	351.740	&	 26.578	&	55424.026	&	0.8035	&	MP	\\
GRB100814A	&	 22.473	&	-17.995	&	55422.160	&	1.44	&	MP	\\
GRB100805A	&	299.876	&	 52.628	&	55413.186	&	1.85	&	\cite{Oates:2012hs}	\\
GRB100728B	&	 44.056	&	  0.281	&	55405.439	&	2.106	&	MP	\\
GRB100728A	&	 88.758	&	-15.256	&	55405.095	&	1.567	&	MP	\\
GRB100724A	&	194.543	&	-11.103	&	55401.029	&	1.288	&	MP	\\
GRB100704A	&	133.642	&	-24.203	&	55381.149	&	3.6    	&	\cite{2010GCN.10940....1C}	\\
GRB100628A	&	225.973	&	-31.664	&	55375.345	&	0.102	&	\cite{2010GCN.10946....1C,NicuesaGuelbenzu}	\\
GRB100625A	&	 15.796	&	-39.088	&	55372.773	&	0.452	&	MP	\\
GRB100621A	&	315.305	&	-51.106	&	55368.127	&	0.542	&	MP	\\
GRB100615A	&	177.205	&	-19.481	&	55362.083	&	1.398	&	MP	\\
GRB100606A	&	350.627	&	-66.241	&	55353.800	&	1.5545	&	\cite{Kruhler}	\\
GRB100518A	&	304.789	&	-24.554	&	55334.482	&	4      	&	MP	\\
GRB100513A	&	169.612	&	  3.628	&	55329.088	&	4.772	&	MP	\\
GRB100508A	&	 76.246	&	-20.711	&	55324.389	&	0.5201	&	MP	\\
GRB100425A	&	299.196	&	-26.431	&	55311.119	&	1.755	&	MP	\\
GRB100424A	&	209.448	&	  1.538	&	55310.689	&	2.465	&	MP	\\
GRB100418A	&	256.362	&	 11.462	&	55304.882	&	0.6235	&	MP	\\
GRB100414A	&	192.112	&	  8.693	&	55300.097	&	1.368	&	MP	\\
GRB100413A	&	266.221	&	 15.834	&	55299.732	&	3.9    	&	\cite{2010GCN.10588....1C}	\\
GRB100316D	&	107.628	&	-56.256	&	55271.531	&	0.059	&	MP	\\
GRB100316B	&	163.488	&	-45.473	&	55271.334	&	1.18	&	MP	\\
GRB100316A	&	251.978	&	 71.827	&	55271.099	&	3.155	&	MP	\\
GRB100302A	&	195.516	&	 74.590	&	55257.829	&	4.813	&	MP	\\
GRB100219A	&	154.202	&	-12.566	&	55246.636	&	4.6667	&	MP	\\
GRB100216A	&	154.251	&	 35.522	&	55243.422	&	0.038	&	\cite{2010GCN.10428....1C,Dichiara}	\\
GRB100213B	&	124.282	&	 43.448	&	55240.957	&	0.604	&	\cite{2010GCN.10422....1C,2010GCN.10913....1L}	\\
GRB100206A	&	 47.163	&	 13.157	&	55233.563	&	0.4068	&	MP	\\
GRB100117A	&	 11.269	&	 -1.595	&	55213.879	&	0.92	&	MP	\\
GRB091208B	&	 29.392	&	 16.890	&	55173.410	&	1.063	&	MP	\\
GRB091127	&	 36.583	&	-18.952	&	55162.976	&	0.49	&	MP	\\
GRB091109	&	309.258	&	-44.158	&	55144.207	&	3.076	&	MP	\\
GRB091029	&	 60.178	&	-55.956	&	55133.162	&	2.752	&	MP	\\
GRB091024	&	339.248	&	 56.890	&	55128.372	&	1.092	&	MP	\\
GRB091020	&	175.730	&	 50.978	&	55124.900	&	1.71	&	MP	\\
GRB091018	&	 32.186	&	-57.548	&	55122.867	&	0.971	&	MP	\\
GRB091003	&	251.520	&	 36.625	&	55107.191	&	0.8969	&	MP	\\
GRB090927	&	343.972	&	-70.980	&	55101.422	&	1.37	&	MP	\\
GRB090926B	&	 46.308	&	-39.006	&	55100.914	&	1.24	&	MP	\\
GRB090926	&	353.400	&	-66.324	&	55100.181	&	2.1062	&	MP	\\
GRB090902B	&	264.939	&	 27.324	&	55076.462	&	1.822	&	MP	\\
GRB090814	&	239.610	&	 25.631	&	55057.036	&	0.696	&	MP	\\
GRB090812	&	353.202	&	-10.605	&	55055.251	&	2.452	&	MP	\\
GRB090809	&	328.680	&	 -0.084	&	55052.730	&	2.737	&	MP	\\
GRB090726	&	248.680	&	 72.884	&	55038.946	&	2.71	&	MP	\\
GRB090715B	&	251.340	&	 44.839	&	55027.877	&	3      	&	MP	\\
GRB090709	&	289.927	&	 60.728	&	55021.319	&	1.8 	&	\cite{Perley:2013fh}	\\
GRB090618	&	293.994	&	 78.357	&	55000.353	&	0.54	&	MP	\\
GRB090530	&	179.419	&	 26.594	&	54981.138	&	1.266	&	MP	\\
GRB090529	&	212.469	&	 24.459	&	54980.592	&	2.625	&	MP	\\
GRB090519	&	142.279	&	  0.180	&	54970.881	&	3.85	&	MP	\\
GRB090516	&	138.260	&	-11.854	&	54967.352	&	4.109	&	MP	\\
GRB090515	&	164.152	&	 14.441	&	54966.198	&	0.403	&	\cite{Berger:2010ag,Fong:2013iia}	\\
GRB090510	&	333.552	&	-26.583	&	54961.016	&	0.903	&	MP	\\
GRB090429B	&	210.667	&	 32.170	&	54950.229	&	9.4    	&	MP	\\
GRB090426	&	189.075	&	 32.986	&	54947.534	&	2.609	&	MP	\\
GRB090424	&	189.521	&	 16.838	&	54945.589	&	0.544	&	MP	\\
GRB090423	&	148.889	&	 18.149	&	54944.330	&	8.26	&	MP	\\
GRB090418	&	269.313	&	 33.406	&	54939.464	&	1.608	&	MP	\\
GRB090417B	&	209.694	&	 47.018	&	54938.639	&	0.345	&	MP	\\
GRB090407	&	 68.980	&	-12.679	&	54928.436	&	1.4485	&	MP	\\
GRB090404	&	239.240	&	 35.516	&	54925.664	&	3      	&	MP	\\
GRB090401B	&	 95.088	&	 -8.972	&	54922.358	&	3.1    	&	\cite{Oates:2012hs}	\\
GRB090328	&	 90.665	&	-41.882	&	54918.401	&	0.736	&	MP	\\
GRB090323	&	190.710	&	 17.053	&	54913.002	&	3.57	&	MP	\\
GRB090313	&	198.401	&	  8.097	&	54903.379	&	3.375	&	MP	\\
GRB090205	&	220.911	&	-27.853	&	54867.961	&	4.6497	&	MP	\\
GRB090201	&	 92.052	&	-46.590	&	54863.741	&	2.1 	&	MP	\\
GRB090113	&	 32.057	&	 33.428	&	54844.778	&	1.7493	&	MP	\\
GRB090102	&	128.244	&	 33.114	&	54833.122	&	1.547	&	MP	\\
GRB081230	&	 37.331	&	-25.148	&	54830.858	&	2   	&	MP	\\
GRB081228	&	 39.462	&	 30.853	&	54828.054	&	3.44	&	MP	\\
GRB081222	&	 22.740	&	-34.095	&	54822.204	&	2.77	&	MP	\\
GRB081221	&	 15.793	&	-24.548	&	54821.681	&	2.26	&	MP	\\
GRB081211B	&	168.265	&	 53.830	&	54811.260	&	0.216	&	\cite{GCN8914}	\\
GRB081210	&	 70.484	&	-11.257	&	54810.847	&	2.0631	&	MP	\\
GRB081203	&	233.032	&	 63.521	&	54803.577	&	2.05	&	MP	\\
GRB081121	&	 89.276	&	-60.603	&	54791.858	&	2.512	&	MP	\\
GRB081118	&	 82.592	&	-43.301	&	54788.623	&	2.58	&	MP	\\
GRB081109	&	330.790	&	-54.711	&	54779.293	&	0.9787	&	MP	\\
GRB081029	&	346.772	&	-68.156	&	54768.072	&	3.8479	&	MP	\\
GRB081028	&	121.895	&	  2.308	&	54767.017	&	3.038	&	MP	\\
GRB081008	&	279.958	&	-57.431	&	54747.832	&	1.9685	&	MP	\\
GRB081007	&	339.960	&	-40.147	&	54746.224	&	0.5295	&	MP	\\
GRB080928	&	 95.070	&	-55.200	&	54737.626	&	1.692	&	MP	\\
GRB080916C	&	119.847	&	-56.638	&	54725.009	&	4.35	&	MP	\\
GRB080916	&	336.276	&	-57.023	&	54725.406	&	0.689	&	MP	\\
GRB080913	&	 65.728	&	-25.130	&	54722.283	&	6.695	&	MP	\\
GRB080906	&	228.044	&	-80.518	&	54715.565	&	2.1 	&	MP	\\
GRB080905B	&	301.741	&	-62.563	&	54714.705	&	2.374	&	MP	\\
GRB080905	&	287.674	&	-18.880	&	54714.499	&	0.1218	&	MP	\\
GRB080825B	&	209.201	&	-68.955	&	54703.741	&	4.3    	&	MP	\\
GRB080810	&	356.793	&	  0.320	&	54688.549	&	3.35	&	MP	\\
GRB080805	&	314.223	&	-62.445	&	54683.321	&	1.505	&	MP	\\
GRB080804	&	328.668	&	-53.185	&	54682.972	&	2.2045	&	MP	\\
GRB080721	&	224.483	&	-11.724	&	54668.434	&	2.591	&	MP	\\
GRB080710	&	  8.274	&	 19.502	&	54657.301	&	0.845	&	MP	\\
GRB080707	&	 32.618	&	 33.109	&	54654.353	&	1.23	&	MP	\\
GRB080607	&	194.947	&	 15.920	&	54624.255	&	3.036	&	MP	\\
GRB080605	&	262.125	&	  4.016	&	54622.992	&	1.6398	&	MP	\\
GRB080604	&	236.965	&	 20.558	&	54621.310	&	1.416	&	MP	\\
GRB080603B	&	176.532	&	 68.061	&	54620.818	&	2.69	&	MP	\\
GRB080603	&	279.409	&	 62.744	&	54620.471	&	1.688	&	MP	\\
GRB080602	&	 19.176	&	 -9.232	&	54619.063	&	1.8204	&	MP	\\
GRB080520	&	280.193	&	-54.992	&	54606.931	&	1.545	&	MP	\\
GRB080517	&	102.242	&	 50.735	&	54603.891	&	0.089	&	MP	\\
GRB080516	&	120.642	&	-26.160	&	54602.012	&	3.2 	&	\cite{GCN7747}	\\
GRB080515	&	  3.163	&	 32.578	&	54601.251	&	2.47	&	MP	\\
GRB080514B	&	322.845	&	  0.708	&	54600.414	&	1.8    	&	MP	\\
GRB080430	&	165.311	&	 51.686	&	54586.828	&	0.767	&	MP	\\
GRB080413B	&	326.144	&	-19.981	&	54569.369	&	1.1    	&	MP	\\
GRB080413	&	287.299	&	-27.678	&	54569.121	&	2.433	&	MP	\\
GRB080411	&	 37.980	&	-71.302	&	54567.886	&	1.03	&	MP	\\
GRB080330	&	169.269	&	 30.623	&	54555.154	&	1.51	&	MP	\\
GRB080325	&	277.893	&	 36.524	&	54550.173	&	1.78	&	MP	\\
GRB080319C	&	258.981	&	 55.392	&	54544.518	&	1.95	&	MP	\\
GRB080319B	&	217.921	&	 36.302	&	54544.259	&	0.937	&	MP	\\
GRB080319	&	206.333	&	 44.080	&	54544.240	&	2.0265	&	MP	\\
GRB080310	&	220.058	&	 -0.175	&	54535.360	&	2.42	&	MP	\\
GRB080210	&	251.267	&	 13.827	&	54506.326	&	2.641	&	MP	\\
GRB080207	&	207.512	&	  7.502	&	54503.896	&	2.0858	&	MP	\\
GRB080205	&	 98.253	&	 62.792	&	54501.330	&	2.72	&	MP	\\
GRB080129	&	105.284	&	 -7.846	&	54494.255	&	4.349	&	MP	\\
GRB080123	&	338.943	&	-64.901	&	54488.182	&	0.495	&	MP	\\
GRB071227	&	 58.130	&	-55.984	&	54461.843	&	0.383	&	MP	\\
GRB071122	&	276.605	&	 47.075	&	54426.058	&	1.14	&	MP	\\
GRB071117	&	335.044	&	-63.443	&	54421.618	&	1.331	&	MP	\\
GRB071112C	&	 39.212	&	 28.371	&	54416.773	&	0.823	&	MP	\\
GRB071031	&	  6.405	&	-58.060	&	54404.046	&	2.692	&	MP	\\
GRB071028B	&	354.163	&	-31.621	&	54401.114	&	0.94	&	\cite{Clemens:2011wi}	\\
GRB071025	&	355.071	&	 31.778	&	54398.173	&	5.2 	&	MP	\\
GRB071021	&	340.643	&	 23.718	&	54394.404	&	2.452	&	MP	\\
GRB071020	&	119.665	&	 32.861	&	54393.293	&	2.145	&	MP	\\
GRB071010B	&	150.539	&	 45.730	&	54383.865	&	0.947	&	MP	\\
GRB071010	&	288.060	&	-32.402	&	54383.154	&	0.98	&	MP	\\
GRB071003	&	301.851	&	 10.947	&	54376.320	&	1.6043	&	MP	\\
GRB070810	&	189.963	&	 10.751	&	54322.092	&	2.17	&	MP	\\
GRB070809	&	203.770	&	-22.142	&	54321.807	&	0.2187	&	MP	\\
GRB070802	&	 36.899	&	-55.528	&	54314.297	&	2.45	&	MP	\\
GRB070724	&	 27.808	&	-18.594	&	54305.454	&	0.457	&	MP	\\
GRB070721B	&	 33.137	&	 -2.195	&	54302.440	&	3.626	&	MP	\\
GRB070714B	&	 57.843	&	 28.298	&	54295.208	&	0.92	&	MP	\\
GRB070714	&	 42.930	&	 30.243	&	54295.139	&	1.58	&	MP	\\
GRB070612	&	121.369	&	 37.269	&	54263.110	&	0.617	&	MP	\\
GRB070611	&	  1.992	&	-29.756	&	54262.081	&	2.04	&	MP	\\
GRB070529	&	283.742	&	 20.659	&	54249.534	&	2.4996	&	MP	\\
GRB070521	&	242.661	&	 30.256	&	54241.286	&	2.0865	&	MP	\\
GRB070518	&	254.199	&	 55.295	&	54238.602	&	1.161	&	MP	\\
GRB070508	&	312.799	&	-78.385	&	54228.179	&	0.82	&	\cite{6398,Xiao:2009dr}	\\
GRB070506	&	347.218	&	 10.722	&	54226.233	&	2.31	&	MP	\\
GRB070429B	&	328.015	&	-38.829	&	54219.131	&	0.904	&	MP	\\
GRB070419B	&	315.707	&	-31.264	&	54209.447	&	1.9591	&	MP	\\
GRB070419	&	182.745	&	 39.925	&	54209.416	&	0.97	&	MP	\\
GRB070411	&	107.333	&	  1.064	&	54201.842	&	2.954	&	MP	\\
GRB070328	&	 65.115	&	-34.067	&	54187.162	&	2.0627	&	MP	\\
GRB070318	&	 48.487	&	-42.946	&	54177.312	&	0.836	&	MP	\\
GRB070306	&	148.097	&	 10.482	&	54165.695	&	1.4965	&	MP	\\
GRB070224	&	179.027	&	-13.330	&	54155.853	&	1.9922	&	MP	\\
GRB070223	&	153.452	&	 43.134	&	54154.052	&	1.6295	&	MP	\\
GRB070208	&	197.886	&	 61.965	&	54139.382	&	1.165	&	MP	\\
GRB070129	&	 37.004	&	 11.684	&	54129.983	&	2.3384	&	MP	\\
GRB070125	&	117.824	&	 31.151	&	54125.306	&	1.547	&	MP	\\
GRB070110	&	  0.913	&	-52.974	&	54110.307	&	2.352	&	MP	\\
GRB070103	&	352.558	&	 26.876	&	54103.866	&	2.6208	&	MP	\\
GRB061222B	&	105.353	&	-25.860	&	54091.174	&	3.355	&	MP	\\
GRB061222	&	358.264	&	 46.533	&	54091.145	&	2.088	&	MP	\\
GRB061217	&	160.413	&	-21.124	&	54086.153	&	0.827	&	MP	\\
GRB061210	&	144.522	&	 15.621	&	54079.514	&	0.4097	&	MP	\\
GRB061202	&	105.525	&	-74.698	&	54071.341	&	2.2543	&	MP	\\
GRB061201	&	332.134	&	-74.580	&	54070.666	&	0.111	&	MP	\\
GRB061126	&	 86.602	&	 64.211	&	54065.367	&	1.1588	&	MP	\\
GRB061121	&	147.227	&	-13.195	&	54060.641	&	1.314	&	MP	\\
GRB061110B	&	323.919	&	  6.876	&	54049.916	&	3.44	&	MP	\\
GRB061110	&	336.291	&	 -2.258	&	54049.491	&	0.758	&	MP	\\
GRB061021	&	145.150	&	-21.952	&	54029.652	&	0.3463	&	MP	\\
GRB061007	&	 46.332	&	-50.501	&	54015.422	&	1.261	&	MP	\\
GRB061006	&	111.031	&	-79.199	&	54014.699	&	0.4377	&	MP	\\
GRB061002	&	220.347	&	 48.742	&	54010.044	&	0.564	&	\cite{Chrimes:2018ptj}	\\
GRB060927	&	329.550	&	  5.364	&	54005.589	&	5.47	&	MP	\\
GRB060926	&	263.932	&	 13.038	&	54004.700	&	3.204	&	MP	\\
GRB060923B	&	238.195	&	-30.904	&	54001.485	&	1.5094	&	MP	\\
GRB060923	&	254.617	&	 12.361	&	54001.217	&	2.6 	&	\cite{Tanvir:2008qy,Perley:2013fh}	\\
GRB060912	&	  5.284	&	 20.972	&	53990.580	&	0.937	&	MP	\\
GRB060908	&	 31.826	&	  0.342	&	53986.373	&	1.8836	&	MP	\\
GRB060906	&	 40.754	&	 30.362	&	53984.356	&	3.6855	&	MP	\\
GRB060904B	&	 58.211	&	 -0.725	&	53982.105	&	0.703	&	MP	\\
GRB060814	&	221.339	&	 20.586	&	53961.960	&	1.9229	&	MP	\\
GRB060805	&	220.931	&	 12.586	&	53952.200	&	2.3633	&	MP	\\
GRB060801	&	213.006	&	 16.982	&	53948.511	&	1.131	&	MP	\\
GRB060729	&	 95.382	&	-62.370	&	53945.800	&	0.54	&	MP	\\
GRB060719	&	 18.432	&	-48.381	&	53935.285	&	1.532	&	MP	\\
GRB060714	&	227.860	&	 -6.566	&	53930.633	&	2.7105	&	MP	\\
GRB060708	&	  7.808	&	-33.759	&	53924.511	&	1.92	&	MP	\\
GRB060707	&	357.080	&	-17.905	&	53923.896	&	3.425	&	MP	\\
GRB060614	&	320.884	&	-53.027	&	53900.530	&	0.125	&	MP	\\
GRB060607	&	329.710	&	-22.496	&	53893.217	&	3.0749	&	MP	\\
GRB060605	&	322.156	&	 -6.059	&	53891.761	&	3.773	&	MP	\\
GRB060604	&	337.229	&	-10.916	&	53890.763	&	2.1357	&	MP	\\
GRB060602	&	149.570	&	  0.304	&	53888.897	&	0.787	&	MP	\\
GRB060526	&	232.826	&	  0.285	&	53881.686	&	3.221	&	MP	\\
GRB060522	&	322.937	&	  2.886	&	53877.091	&	5.11	&	MP	\\
GRB060512	&	195.774	&	 41.191	&	53867.968	&	2.1 	&	MP	\\
GRB060510B	&	239.122	&	 78.570	&	53865.349	&	4.94	&	MP	\\
GRB060510	&	 95.867	&	 -1.163	&	53865.322	&	1.2 	&	\cite{Oates:2012hs}	\\
GRB060505	&	331.764	&	-27.815	&	53860.275	&	0.089	&	MP	\\
GRB060502B	&	278.938	&	 52.632	&	53857.725	&	0.287	&	MP	\\
GRB060502	&	240.927	&	 66.601	&	53857.127	&	1.51	&	MP	\\
GRB060418	&	236.428	&	 -3.639	&	53843.129	&	1.4895	&	MP	\\
GRB060319	&	176.388	&	 60.011	&	53813.039	&	1.172	&	MP	\\
GRB060306	&	 41.095	&	 -2.148	&	53800.034	&	1.559	&	MP	\\
GRB060223	&	 55.206	&	-17.130	&	53789.253	&	4.41	&	MP	\\
GRB060218	&	 50.416	&	 16.867	&	53784.149	&	0.0331	&	MP	\\
GRB060210	&	 57.739	&	 27.026	&	53776.208	&	3.91	&	MP	\\
GRB060206	&	202.931	&	 35.051	&	53772.199	&	4.048	&	MP	\\
GRB060204B	&	211.812	&	 27.677	&	53770.607	&	2.3393	&	MP	\\
GRB060202	&	 35.846	&	 38.384	&	53768.362	&	0.783	&	MP	\\
GRB060124	&	 77.108	&	 69.741	&	53759.663	&	2.296	&	MP	\\
GRB060123	&	179.700	&	 45.514	&	53758.932	&	0.56	&	MP	\\
GRB060121	&	137.466	&	 45.663	&	53756.933	&	4.6 	&	\cite{deUgartePostigo:2006ub}	\\
GRB060116	&	 84.693	&	 -5.437	&	53751.359	&	6.6 	&	\cite{2006GCN..4545....1G}	\\
GRB060115	&	 54.035	&	 17.345	&	53750.547	&	3.53	&	MP	\\
GRB060111	&	276.205	&	 37.604	&	53746.183	&	2.32	&	MP	\\
GRB051227	&	125.241	&	 31.925	&	53731.755	&	0.8 	&	MP	\\
GRB051221	&	328.703	&	 16.891	&	53725.077	&	0.5468	&	MP	\\
GRB051210	&	330.171	&	-57.614	&	53714.241	&	2.5 	&	MP	\\
GRB051117B	&	 85.181	&	-19.274	&	53691.558	&	0.481	&	MP	\\
GRB051111	&	348.138	&	 18.375	&	53685.250	&	1.5495	&	MP	\\
GRB051109B	&	345.460	&	 38.680	&	53683.361	&	0.08	&	MP	\\
GRB051109	&	330.314	&	 40.823	&	53683.050	&	2.346	&	MP	\\
GRB051022	&	359.017	&	 19.607	&	53665.547	&	0.809	&	MP	\\
GRB051016B	&	132.116	&	 13.656	&	53659.770	&	0.9364	&	MP	\\
GRB051008	&	202.873	&	 42.098	&	53651.690	&	2.77	&	MP	\\
GRB051006	&	110.809	&	  9.506	&	53649.855	&	1.059	&	MP	\\
GRB051001	&	350.953	&	-31.523	&	53644.466	&	2.4296	&	MP	\\
GRB050922C	&	317.388	&	 -8.758	&	53635.830	&	2.198	&	MP	\\
GRB050922B	&	  5.806	&	 -5.605	&	53635.626	&	4.5    	&	MP	\\
GRB050915	&	 81.687	&	-28.016	&	53628.474	&	2.5273	&	MP	\\
GRB050911	&	 13.657	&	-38.849	&	53624.666	&	0.165	&	\cite{2005GCN..3962....1B,Berger:2006hs}	\\
GRB050908	&	 20.461	&	-12.955	&	53621.238	&	3.344	&	MP	\\
GRB050904	&	 13.711	&	 14.085	&	53617.078	&	6.29	&	MP	\\
GRB050826	&	 87.756	&	 -2.643	&	53608.263	&	0.297	&	MP	\\
GRB050824	&	 12.234	&	 22.609	&	53606.967	&	0.83	&	MP	\\
GRB050822	&	 51.113	&	-46.033	&	53604.159	&	1.434	&	MP	\\
GRB050820	&	337.409	&	 19.560	&	53602.274	&	2.612	&	MP	\\
GRB050819	&	358.757	&	 24.861	&	53601.683	&	2.5043	&	MP	\\
GRB050814	&	264.189	&	 46.339	&	53596.485	&	5.77	&	MP	\\
GRB050813	&	241.989	&	 11.248	&	53595.281	&	0.72	&	\cite{Prochaska:2005qf,Ferrero:2006be}	\\
GRB050803	&	350.658	&	  5.786	&	53585.801	&	3.5 	&	MP	\\
GRB050802	&	219.274	&	 27.787	&	53584.422	&	1.71	&	MP	\\
GRB050801	&	204.146	&	-21.928	&	53583.769	&	1.38	&	MP	\\
GRB050730	&	212.072	&	 -3.772	&	53581.832	&	3.9678	&	MP	\\
GRB050724	&	246.185	&	-27.541	&	53575.524	&	0.2575	&	MP	\\
GRB050714B	&	169.699	&	-15.547	&	53565.945	&	2.4383	&	MP	\\
GRB050709	&	345.362	&	-38.978	&	53560.942	&	0.16	&	MP	\\
GRB050525	&	278.136	&	 26.339	&	53515.002	&	0.606	&	MP	\\
GRB050509B	&	189.058	&	 28.984	&	53499.167	&	0.2255	&	MP	\\
GRB050505	&	141.764	&	 30.273	&	53495.974	&	4.27	&	MP	\\
GRB050502B	&	142.542	&	 16.997	&	53492.393	&	5.2    	&	MP	\\
GRB050502	&	202.443	&	 42.674	&	53492.093	&	3.793	&	MP	\\
GRB050416	&	188.478	&	 21.057	&	53476.462	&	0.6535	&	MP	\\
GRB050408	&	180.573	&	 10.852	&	53468.683	&	1.2357	&	MP	\\
GRB050406	&	 34.468	&	-50.188	&	53466.666	&	2.44	&	\cite{Schady:2006rn}	\\
GRB050401	&	247.870	&	  2.187	&	53461.597	&	2.9 	&	MP	\\
GRB050319	&	154.200	&	 43.548	&	53448.397	&	3.24	&	MP	\\
GRB050318	&	 49.713	&	-46.396	&	53447.656	&	1.44	&	MP	\\
GRB050315	&	306.476	&	-42.600	&	53444.875	&	1.949	&	MP	\\
GRB050223	&	271.385	&	-62.472	&	53424.131	&	0.5915	&	MP	\\
GRB050219	&	166.412	&	-40.683	&	53420.528	&	0.211	&	MP	\\
GRB050215B	&	174.448	&	 40.797	&	53416.107	&	2.62	&	MP	\\
GRB050126	&	278.113	&	 42.370	&	53396.501	&	1.29	&	MP	\\
GRB041006	&	 13.709	&	  1.235	&	53284.512	&	0.716	&	MP	\\
GRB040924	&	 31.594	&	 16.114	&	53272.494	&	0.859	&	MP	\\
GRB040912	&	359.226	&	 -1.001	&	53260.592	&	1.563	&	MP	\\
GRB040701	&	311.943	&	-40.237	&	53189.583	&	0.2146	&	\cite{2004GCN..2630....1F,Soderberg:2005ct}	\\
GRB031203	&	120.626	&	-39.850	&	52976.918	&	0.105	&	MP	\\
GRB030528	&	256.008	&	-22.650	&	52787.544	&	0.782	&	MP	\\
GRB030429	&	183.281	&	-20.914	&	52758.446	&	2.656	&	MP	\\
GRB030329	&	161.208	&	 21.522	&	52727.484	&	0.1685	&	MP	\\
GRB030328	&	182.702	&	 -9.348	&	52726.473	&	1.52	&	MP	\\
GRB030323	&	166.539	&	-21.770	&	52721.915	&	3.372	&	MP	\\
GRB030226	&	173.271	&	 25.898	&	52696.157	&	1.986	&	MP	\\
GRB030115	&	169.625	&	 15.001	&	52654.141	&	2.5 	&	\cite{Levan:2006pd}	\\
GRB021211	&	122.270	&	  6.678	&	52619.471	&	1.004	&	MP	\\
GRB021004	&	  6.745	&	 18.949	&	52551.504	&	2.33	&	MP	\\
GRB020903	&	342.176	&	-20.769	&	52520.421	&	0.25	&	MP	\\
GRB020819B	&	351.840	&	  6.250	&	52505.623	&	1.9621	&	MP	\\
GRB020813	&	296.658	&	-19.588	&	52499.114	&	1.2545	&	MP	\\
GRB020405	&	209.513	&	-31.373	&	52369.029	&	0.69	&	MP	\\
GRB020305	&	190.616	&	-14.303	&	52338.629	&	0.2 	&	MP	\\
GRB020127	&	123.756	&	 36.776	&	52301.873	&	1.9 	&	MP	\\
GRB020124	&	143.211	&	-11.520	&	52298.445	&	3.198	&	MP	\\
GRB011211	&	168.825	&	-21.949	&	52254.798	&	2.14	&	MP	\\
GRB011121	&	173.606	&	-76.028	&	52234.783	&	0.36	&	MP	\\
GRB010921	&	344.000	&	 40.931	&	52173.219	&	0.45	&	MP	\\
GRB010222	&	223.052	&	 43.018	&	51962.308	&	1.477	&	MP	\\
GRB000926	&	256.040	&	 51.786	&	51813.877	&	2.066	&	MP	\\
GRB000911	&	 34.681	&	  7.741	&	51798.302	&	1.0585	&	MP	\\
GRB000607	&	 38.496	&	 17.002	&	51702.101	&	0.1405	&	\cite{Gal-Yam:2005xjy}	\\
GRB000418	&	186.330	&	 20.103	&	51652.896	&	1.118	&	MP	\\
GRB000301C	&	245.078	&	 29.443	&	51604.529	&	2.03	&	MP	\\
GRB000214	&	283.613	&	-66.458	&	51588.042	&	0.42	&	MP	\\
GRB000210	&	 29.815	&	-40.659	&	51584.364	&	0.8463	&	MP	\\
GRB000131	&	 93.379	&	-51.944	&	51574.624	&	4.5 	&	MP	\\
GRB991216	&	 77.380	&	 11.285	&	51528.672	&	1.02	&	MP	\\
GRB991208	&	248.473	&	 46.456	&	51520.192	&	0.706	&	MP	\\
GRB990712	&	337.971	&	-73.408	&	51371.323	&	0.434	&	MP	\\
GRB990705	&	 77.477	&	-72.131	&	51364.668	&	0.842	&	MP	\\
GRB990510	&	204.525	&	-80.500	&	51308.367	&	1.619	&	MP	\\
GRB990506	&	178.709	&	-26.676	&	51304.475	&	1.3 	&	MP	\\
GRB990123	&	231.374	&	 44.758	&	51201.408	&	1.6 	&	MP	\\
GRB981226	&	352.405	&	-23.932	&	51173.408	&	1.11	&	\cite{Christensen:2005te}	\\
GRB980703	&	359.779	&	  8.585	&	50997.182	&	0.966	&	MP	\\
GRB980613	&	154.442	&	 71.486	&	50977.202	&	1.096	&	MP	\\
GRB980425	&	293.725	&	-52.819	&	50928.446	&	0.0085	&	MP	\\
GRB980329	&	105.658	&	 38.846	&	50901.156	&	3.5 	&	MP	\\
GRB980326	&	129.143	&	-18.857	&	50898.888	&	1       &	\cite{Bloom:1999aa}	\\
GRB971214	&	179.110	&	 65.200	&	50796.110	&	3.42	&	MP	\\
GRB970828	&	272.106	&	 59.302	&	50688.739	&	0.9578	&	MP	\\
GRB970508	&	103.456	&	 79.272	&	50576.904	&	0.835	&	MP	\\
GRB970228	&	 75.488	&	 11.768	&	50507.124	&	0.695	&	MP	\\
\hline
\end{longtable}

\section{GRBs relevant for Figure 1}\label{sec:GRBneutprop}

In this appendix we describe some properties of the GRB-neutrino events appearing in Fig.~\ref{fig:1}. 

As already stressed in the main text, our analysis relied  on the latest HESE neutrino data release by IceCube~\cite{icecubedatarelease}.
We considered only shower neutrinos, for which the energy is approximately equal to the visible energy~\cite{IceCubeEnergy2013,jackJrEnergy2018} (whereas the energy of track neutrinos is poorly known).

The GRBs which we analyzed (also in producing simulated data) are those of our catalogue reported
in Appendix~\ref{GRBcatalogue}. It is worth emphasizing that our catalogue agrees with both~\cite{icecubeGRBs}
and~\cite{maxplanckGRBs} for what concerns the four GRBs relevant for Fig.~\ref{fig:1}. 

The  neutrino with the lowest energy among those relevant for Figure~\ref{fig:1} has a visible energy of 55.3~TeV and is associated to GRB110503A~\cite{GRB110503Adetection}, a long GRB ($\mathrm{T}90=10.0$~s) with measured redshift of 1.613~\cite{GRB110503Aredshift}.

The violet point in Fig.~\ref{fig:1}, which is the one that was already taken into account in the analysis of Ref.~\cite{natastro2023}, has a visible energy of 64.0~TeV, and is associated to GRB111229A~\cite{GRB111229Adetection}, a long GRB ($\mathrm{T}90=25.4$~s)
with measured redshift of 1.38~\cite{GRB111229Aredshift}. 

The neutrino with visible energy of 302.2~TeV relevant for Fig.~1 is associated to GRB120923A~\cite{GRB120923Adetection}, a long GRB ($\mathrm{T}90=27.2$~s) with measured redshift of 7.8~\cite{GRB120923Aredshift}.

The highest-energy neutrino relevant for Fig.~1 (also known as ``big bird"~\cite{BigBird}) has visible energy of 2075.0~TeV 
and is associated with GRB120909A~\cite{GRB120909Adetection}, a long GRB ($\mathrm{T}90=112.1$~s) with measured redshift of 3.93~\cite{GRB120909Aredshift}.

\end{document}